\documentclass[11pt,a4paper,english]{article}
\usepackage[utf8]{inputenc} 
\usepackage[T1]{fontenc}    
\usepackage{babel}
\usepackage{blindtext}
\usepackage{hyperref}       
\usepackage{url}            
\usepackage{booktabs}       
\usepackage{amsfonts}       
\usepackage{nicefrac}       
\usepackage{microtype}      
\usepackage{xcolor}         
\usepackage{amsmath,amssymb}
\usepackage{wrapfig}
\usepackage{stfloats}
\usepackage{graphicx}
\usepackage{placeins}  
\usepackage{float}    
\usepackage{caption}
\usepackage{subcaption}
\usepackage{bm}
\usepackage{listings}
\usepackage{multicol}
\usepackage{multirow} 
\usepackage{pgffor}
\usepackage{cleveref}
\usepackage{geometry}

\NewDocumentCommand{\codeword}{v}{%
\texttt{\textcolor{blue}{#1}}%
}

\title{Recurrent Neural Networks for Dynamic VWAP Execution: Adaptive Trading Strategies with Temporal Kolmogorov-Arnold Networks}

\author{%
    Rémi Genet \\
    \small DRM, Université Paris Dauphine - PSL \\
    \small Aplo \\
    \small remi.genet@dauphine.psl.eu \\
}

\begin{document}

\maketitle

\begin{abstract}
The execution of Volume Weighted Average Price (VWAP) orders remains a critical challenge in modern financial markets, particularly as trading volumes and market complexity continue to increase. In my previous work \cite{genet2025staticvwap}, I introduced a novel deep learning approach that demonstrated significant improvements over traditional VWAP execution methods by directly optimizing the execution problem rather than relying on volume curve predictions. However, that model was static because it employed the fully linear approach described in \cite{genet2024tln}, which is not designed for dynamic adjustment. This paper extends that foundation by developing a dynamic neural VWAP framework that adapts to evolving market conditions in real time. We introduce two key innovations: first, the integration of recurrent neural networks to capture complex temporal dependencies in market dynamics, and second, a sophisticated dynamic adjustment mechanism that continuously optimizes execution decisions based on market feedback. The empirical analysis, conducted across five major cryptocurrency markets, demonstrates that this dynamic approach achieves substantial improvements over both traditional methods and our previous static implementation, with execution performance gains of 10–15\% in liquid markets and consistent outperformance across varying conditions. These results suggest that adaptive neural architectures can effectively address the challenges of modern VWAP execution while maintaining computational efficiency suitable for practical deployment.
\end{abstract}

\newpage

\section{Introduction}
The execution of large trading orders in financial markets is a complex and strategic challenge, with the potential to significantly impact both transaction costs and overall market dynamics. In this context, the concept of Volume Weighted Average Price (VWAP) has emerged as a key benchmark and execution strategy, offering market participants a robust framework for minimizing market impact while closely tracking average traded prices over a specified period of time. The practical and theoretical importance of VWAP execution has grown with the rapid evolution of electronic and algorithmic trading platforms. As documented by Mackenzie \cite{Mackenzie}, algorithmic trading now accounts for a substantial majority of institutional order flow. VWAP strategies represent a key component of this automated execution landscape. Despite this practical prominence, however, the academic literature has historically placed greater emphasis on alternative execution benchmarks such as Implementation Shortfall (IS) \cite{perold1988implementation}, opening a door for research in the design and performance of VWAP strategies. Initially, VWAP execution aims to address two fundamental objectives. First, by distributing an order over time and closely tracking the average traded price, VWAP strategies seek to minimize the market impact of large trades - a key component of overall transaction costs as established by Berkowitz et al. \cite{TotalCostOfTransactions}. Second, by targeting a pre-defined benchmark, VWAP provides a transparent and objective measure of the execution quality. This metric is crucial  for institutional investors who must justify their trading performance to stakeholders.

\subsection{VWAP Definition and Discretization}

Following the seminal work of Konishi \cite{Konishi}, the Volume Weighted Average Price (VWAP) over a time period \([0,T]\) is defined as:
\begin{equation}
    \text{VWAP}_{[0,T]} = \frac{\int_0^T P(t)V(t)\,dt}{\int_0^T V(t)\,dt},
\end{equation}
where \(P(t)\) and \(V(t)\) denote the price and volume at time \(t\), respectively. As noted by McCulloch and Kazakov \cite{Culoch2007}, financial markets operate in discrete time intervals, which leads to the discretized form:
\begin{equation}
    \text{VWAP}_{[0,T]} = \frac{\sum_{t=1}^{T} P_t\,V_t}{\sum_{t=1}^{T} V_t},
\end{equation}
where \(P_t\) and \(V_t\) represent the price and volume in the \(t\)-th time interval.

For a trader executing a large order of total size \(Q\), Humphery-Jenner \cite{Humphery} shows that the objective is to minimize the difference between the achieved execution price and the market VWAP. Let \(q_t\) denote the quantity traded in interval \(t\), so that:
\begin{equation}
\sum_{t=1}^{T} q_t = Q.
\end{equation}
The execution price is then given by:
\begin{equation}
P_{\text{exec}} = \frac{\sum_{t=1}^{T} P_t\,q_t}{Q}.
\end{equation}

Following Bialkowski et al. \cite{LeFol2006}, the VWAP execution problem can be formulated as minimizing the slippage:
\begin{equation}
\min_{q_1,\ldots,q_T} \left|\frac{\sum_{t=1}^{T} P_t\,q_t}{Q} - \frac{\sum_{t=1}^{T} P_t\,V_t}{\sum_{t=1}^{T} V_t}\right|.
\end{equation}

For clarity, the normalized order allocation is defined as \(\tilde{q}_t = \frac{q_t}{Q}\) (so that \(\sum_{t=1}^{T} \tilde{q}_t = 1\)) and the normalized market volume profile as \(\tilde{V}_t = \frac{V_t}{\sum_{t=1}^{T} V_t}\) (with \(\sum_{t=1}^{T} \tilde{V}_t = 1\)). With these definitions, the execution price becomes:
\begin{equation}
P_{\text{exec}} = \sum_{t=1}^{T} P_t\,\tilde{q}_t,
\end{equation}
and the market VWAP is:
\begin{equation}
\text{VWAP} = \sum_{t=1}^{T} P_t\,\tilde{V}_t.
\end{equation}

Genet \cite{genet2025staticvwap} reformulates the slippage as a bound:\begin{equation}
    S_T \le \sum_{t=1}^{T} \left| \bigl(P_t-\text{VWAP}_t\bigr)\tilde{q}_t \right| + \sum_{t=1}^{T} \left| \text{VWAP}_t\Bigl(\tilde{q}_t-\tilde{V}_t\Bigr) \right|,
\end{equation}
where \(\text{VWAP}_t\) denotes the market VWAP computed over interval \(t\). The first term quantifies the impact of price deviations weighted by the trader's participation rate, while the second term captures the error due to discrepancies between the trader's normalized allocation \(\tilde{q}_t\) and the market's volume fraction \(\tilde{V}_t\). Under the volume conservation constraint and non-negativity constraints \(\tilde{q}_t \geq 0\), this decomposition separates the overall slippage \(S_T\) into a price deviation component and a volume allocation error component. This optimization problem is particularly challenging because future prices and volumes are unknown at execution time, which necessitates accurate predictions of market dynamics while managing execution risk \cite{frei}. Furthermore, the increasing sophistication of market participants and the rising prominence of transaction cost analysis (TCA) in institutional trading \cite{Madhavan2002} have amplified the importance of VWAP execution strategies. As markets evolve, executing large orders while minimizing market impact becomes increasingly complex, thereby requiring more advanced approaches to VWAP execution.

\subsection{Classical VWAP Approaches}

The theoretical foundation of VWAP execution strategies emerged from a series of seminal works that established the mathematical framework for optimal order execution. Konishi \cite{Konishi} provided one of the first comprehensive analyses of VWAP strategies, demonstrating that in markets where volume and volatility are uncorrelated, the optimal execution curve mirrors the expected relative market volume curve. He further extended his analysis to cover the case where volume and volatility are correlated. This breakthrough provided mathematical validation for practitioners' empirical observations and established a theoretical benchmark for subsequent research in the field. Building on this foundation, McCulloch and Kazakov \cite{Culoch2007} developed a more sophisticated model that incorporated practical constraints and information asymmetries. Their work introduced two crucial elements: constrained trading rates and potential information advantages, acknowledging that traders or brokers might possess sensitive information that could influence their attempts to outperform the VWAP benchmark. Their research also revealed important stylized facts about expected relative volume patterns—most notably, the characteristic S-shape observed in equity markets and the finding that higher-turnover stocks exhibit less variation in their expected relative volume. These patterns echoed the well-documented "U" shape effect in equity market trading activity, thereby providing a crucial link between microstructure theory and practical execution strategies.

\medskip

The relationship between volumes and other market variables has been extensively studied in the literature. Notable contributions include the work of Easley and O'Hara \cite{Easley1987}, Viswanathan and Foster \cite{Foster}, Tauchen and Pitts \cite{Tauchen1983}, and Karpoff \cite{Karpoff1987}, who examined volumes as covariates in analyzing and explaining target variables such as price and volatility, though primarily focusing on low-frequency data rather than intraday patterns. Gourieroux et al. \cite{Gourieroux} made significant contributions to the measurement of market trading activity, providing a theoretical framework for understanding trading volume dynamics. McCulloch and Kazakov \cite{Culoch2012} further extended this line of research by transforming Konishi's fixed model into a continuous dynamic framework. This work established the crucial connection between optimal VWAP trading strategies and accurate intraday volume estimation, demonstrating that successful execution depends fundamentally on the ability to anticipate and adapt to volume patterns throughout the trading day.

\medskip

The evolution of these classical approaches reflects a growing recognition of the complexity inherent in VWAP execution. While these models provided valuable insights and theoretical foundations, they also revealed the limitations of purely static approaches in capturing the dynamic nature of modern markets. This recognition would eventually lead to the development of more sophisticated dynamic approaches and, ultimately, to the application of machine learning techniques in VWAP execution strategies.

\subsection{Dynamic Volume Approaches}

A significant paradigm shift in VWAP execution strategies occurred with the introduction of dynamic volume estimation approaches. Bialkowski et al. \cite{LeFol2006} pioneered this advancement by proposing a novel method for estimating intraday volumes through component decomposition. Their work, later refined in Bialkowski et al. \cite{LeFol2012}, separated volume patterns into two distinct components: one reflecting broader market evolution and another capturing stock-specific patterns. This decomposition enabled more accurate volume predictions by modeling the dynamic component using ARMA and SETAR models, demonstrating substantially improved accuracy compared to traditional static approaches. However, transitioning from simplistic volume modeling to these more advanced methods comes at a cost: such approaches no longer explicitly account for the volume–volatility relationship, as it becomes much more challenging to realistically incorporate both components simultaneously. The shift from static to dynamic approaches was further advanced by Humphery-Jenner \cite{Humphery}, who introduced the concept of Dynamic VWAP (DVWAP) in contrast to the traditional Historical VWAP (HVWAP). Their research highlighted a crucial limitation of historical approaches—their inability to incorporate real-time market information during execution. By developing a framework that adapts to incoming news and market developments, they demonstrated significant improvements over historical methods in both basic VWAP tracking and the management of market dynamics.

\medskip

Alternative theoretical perspectives emerged through the work of Bouchard and Dang \cite{bouchard} and Frei and Westray \cite{frei}, who approached VWAP execution through the lens of stochastic analysis. As Frei and Westray \cite{frei} noted, their derived optimal trading rates depended primarily on volume curves rather than price processes, reflecting the assumption of uncorrelated Brownian motion in price movements. This theoretical framework provided valuable insights into the relationship between volume patterns and execution strategy, even as it highlighted the limitations of purely stochastic approaches. A significant contribution to the practical aspects of VWAP execution came from Carmona and Li \cite{Tianhui}, who examined the strategic considerations at both macro and micro scales. Their research was particularly notable for addressing the practical dilemma faced by brokers in choosing between aggressive and passive orders at the high-frequency level, bringing theoretical insights to bear on practical execution decisions. Guéant and Royer \cite{Gueant} made two crucial contributions that addressed previously understudied aspects of VWAP execution. First, they incorporated a comprehensive market impact model that considered both temporary and permanent effects, addressing a critical concern for institutional investors using VWAP orders to manage large positions. Second, they developed a framework for pricing guaranteed VWAP services using CARA utility functions and indifference pricing. This work represented a significant shift from traditional approaches focused solely on benchmark tracking, introducing a more nuanced understanding of risk-adjusted optimal execution. These dynamic approaches collectively highlighted a crucial insight: while modeling market volumes is important, the assumption of independence between prices and volumes often fails to reflect market reality. This recognition, combined with the increasing availability of computational power and market data, set the stage for the application of more sophisticated analytical techniques, particularly in the domain of machine learning and artificial intelligence.

\subsection{The Rise of Deep Learning in Financial Time Series}
In parallel with these theoretical advances, the field of machine learning has witnessed a rapid development of powerful techniques and architectures, particularly in the domain of deep learning. The field of time series analysis and prediction has been fundamentally transformed by developments in deep learning, particularly in the domain of neural networks. As documented by Sezer et al. \cite{sezer2020financial} in their comprehensive review, deep learning models have increasingly outperformed traditional machine learning approaches across various financial forecasting tasks. The evolution of deep learning architectures for financial applications has been marked by several key innovations. The introduction of Long Short-Term Memory (LSTM) networks by Hochreiter and Schmidhuber \cite{hochreiter1997} addressed the vanishing gradient problem that had limited traditional recurrent neural networks, enabling effective learning of long-term dependencies in sequential data. This was followed by the development of Gated Recurrent Units (GRU) by Cho et al. \cite{cho2014}, offering comparable performance with a more streamlined architecture. A revolutionary step forward came with the introduction of attention mechanisms Bahdanau et al. \cite{bahdanau2014neural}, culminating in the Transformer architecture Vaswani et al. \cite{vaswani2017attention}. While initially developed for natural language processing, these architectures' ability to capture both local and global dependencies in sequential data made them particularly suitable for financial time series analysis.

\medskip

During the last decade, we have seen an explosion of deep learning applications in finance, with researchers tackling increasingly complex challenges. Ackerer et al. \cite{ackerer2020deep} demonstrated the power of neural networks in fitting and predicting implied volatility surfaces, while Horvath et al. \cite{horvath2019deep} showed how deep learning could revolutionize pricing and calibration in volatility models. As highlighted by Zhang et al. \cite{zhang2023deep} in their recent review, deep learning models are gradually replacing traditional statistical and machine learning models as the preferred choice for price forecasting tasks. In the specific domain of trading volume prediction, significant advances have been made through the development of specialized architectures such as Temporal Kolmogorov-Arnold Networks (TKAN) \cite{genet2024tkan}, Signature-Weighted Kolmogorov-Arnold Networks (SigKAN) \cite{inzirillo2024sigkan}, Temporal Kolmogorov-Arnold Transformers (TKAT) \cite{genet2024tkat}, Kolmogorov-Arnold Mixture of Experts (KAMoE) \cite{inzirillo2024kamoe} and Recurrent Neural Networks with Signature-Based Gating Mechanisms (SigGate) \cite{genet2025siggate}. 

\subsection{Deep Learning Approaches to Market Execution}

The application of deep learning to market execution problems has evolved significantly in recent years. Early approaches focused primarily on using neural networks for price prediction or simple trading signals. However, the complexity of VWAP execution, with its intricate relationship between volume patterns, price impact, and timing decisions, presents unique challenges that require more sophisticated approaches. Recent research has begun to explore more advanced applications of deep learning to execution problems. Papanicolaou et al. \cite{papanicolaou2023optimal} demonstrated the effectiveness of using LSTMs for large order execution within the Almgren and Chriss framework, showing how deep learning models could capture cross-sectional relationships between different stocks' execution characteristics. While not specifically focused on VWAP execution, this work highlighted the potential for neural networks to learn complex relationships in market impact and execution timing. One particularly promising development has been the recent introduction of Temporal Kolmogorov-Arnold Networks (TKAN) Genet and Inzirillo \cite{genet2024tkan}. This architecture combines the representational power of Kolmogorov-Arnold Networks with sophisticated temporal processing capabilities, demonstrating exceptional performance specifically in cryptocurrency volume prediction tasks. The success of TKANs in volume prediction suggests their potential applicability to the broader challenge of VWAP execution optimization.

\subsection{From Static to Dynamic Neural VWAP}
My previous research Genet \cite{genet2025staticvwap} established a novel approach to VWAP execution by leveraging deep learning techniques in a fundamentally different way from existing methods. Rather than focusing on volume curve prediction like traditional approaches, I demonstrated that directly optimizing the execution strategy through neural networks could significantly improve performance. Using a Temporal Linear Network (TLN), this static approach showed particular effectiveness in handling market uncertainty and extreme events, consistently outperforming conventional methods across various market conditions. However, the inherent limitations of static approaches become particularly apparent in highly volatile markets such as cryptocurrencies, where market conditions can change dramatically within a single execution window. The success of TKANs in cryptocurrency volume prediction, combined with these limitations of static approaches, suggests a natural evolution toward a more dynamic framework. This observation aligns with earlier findings from Bialkowski et al. \cite{LeFol2012} and Humphery-Jenner \cite{Humphery} about the importance of adapting to changing market conditions, but approaches the problem with the enhanced capabilities offered by modern deep learning architectures.

\medskip

In this paper, I propose a dynamic VWAP execution framework that represents a significant advancement in several key aspects:
First, it maintains the robust foundation of our static approach while incorporating adaptive capabilities through recurrent neural networks. This design choice allows our model to preserve the reliable performance characteristics that made the static approach successful while adding the flexibility to adjust execution strategies based on evolving market conditions. Second, our framework leverages the TKAN architecture, which has demonstrated superior performance in volume prediction tasks Genet and Inzirillo \cite{genet2024tkan}. By applying this architecture to execution strategy rather than just volume prediction, I extend its capabilities to address the full complexity of VWAP execution, including market impact considerations and optimal timing decisions. Third, I address a fundamental limitation of existing approaches by directly incorporating market feedback into our execution decisions. This design creates a more responsive system that can adapt to changing market conditions while maintaining the execution constraints necessary for effective VWAP targeting.

\section{Dynamic VWAP Architecture}

The proposed dynamic VWAP execution framework extends previous work on static neural VWAP optimization. By integrating recurrent neural networks and introducing a novel sequential optimization mechanism, the framework adapts to evolving market conditions while preserving the key principles of VWAP benchmarking. This section provides a detailed overview of the model architecture, including its main components, design considerations, and implementation details.

\begin{figure}[H]
    \centering
    \includegraphics[width=\linewidth]{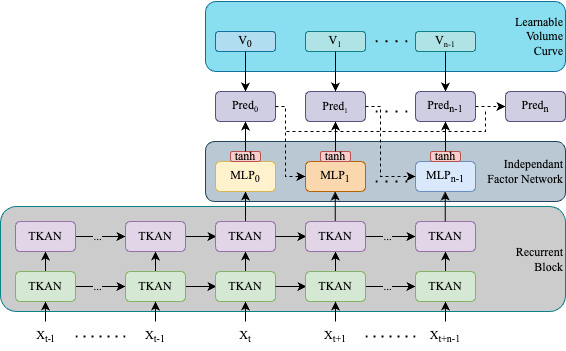}
    \caption{Overview of the dynamic VWAP execution architecture.}
    \label{fig:dynamic_model_overview}
\end{figure}

\subsection{Model Overview}

At its core, the dynamic VWAP framework processes a sequence of market observations to generate optimal execution trajectories in real time. Let \(x_t \in \mathbb{R}^d\) denote the input features at time \(t\), where \(d\) is the feature dimension. Given a lookback period \(l\) and a prediction horizon \(h\), the model operates on input sequences of length \(l + h - 1\), thereby incorporating both historical context and market information received during execution. The architecture consists of three key components: a learnable base volume curve, a recurrent neural network for dynamic adaptation, and a volume adjustment mechanism. These components work together to generate execution trajectories that minimize VWAP slippage while satisfying practical trading constraints.

\subsection{Base Volume Curve}
The foundation of the model is a learnable base volume curve \(v_b \in \mathbb{R}^h\), initialized uniformly as:
\begin{equation}
    v_b^{(0)} = \left\{\frac{1}{h}, \frac{1}{h}, \dots, \frac{1}{h}\right\}.
\end{equation}
This curve is constrained to satisfy two key properties:
\begin{equation}
\sum_{i=1}^h v_b^{(i)} = 1, \quad v_b^{(i)} \geq 0 \quad \forall i \in \{1, \dots, h\}.
\end{equation}

The inclusion of a learnable base curve is motivated by the findings of the static VWAP research Genet \cite{genet2025staticvwap}, which demonstrated the robustness and effectiveness of fixed volume profiles in minimizing execution slippage. By allowing the base curve to adapt during training, the model can potentially discover superior static profiles that serve as the foundation for further dynamic adjustments.

\subsection{Dynamic Volume Adjustment}

The dynamic component begins with a recurrent neural network $\mathcal{R}$ that processes the input sequence to produce hidden states:
\begin{equation}
    h_t = \mathcal{R}(x_t, h_{t-1}),
\end{equation}
where $h_t \in \mathbb{R}^m$ represents the hidden state at time $t$ with dimension $m$.
Note that $h_t$ depends solely on information available up to time $t$, ensuring causality by using both the previous state and the new information at time $t$.

\subsubsection{RNN Architecture Selection}

The recurrent component \(\mathcal{R}\) can be implemented using different RNN variants. Two approaches are considered: Long Short-Term Memory (LSTM) and Temporal Kolmogorov-Arnold Networks (TKAN). LSTM serves as a well-established baseline with proven capability in handling temporal dependencies, while TKAN provides specialized memory management particularly suited for volume forecasting tasks.

\subsection*{LSTM Architecture}

The LSTM unit processes input vector $x_t \in \mathbb{R}^d$ through a series of gates that control information flow. The architecture maintains two types of state: the cell state $c_t$ and the hidden state $h_t$. The key components are the forget gate, which determines what information to discard from the previous state:
\begin{equation}
    f_t = \sigma(W_f x_t + U_f h_{t-1} + b_f),
\end{equation}
The input gate, controlling the integration of new information:
\begin{equation}
    i_t = \sigma(W_i x_t + U_i h_{t-1} + b_i),
\end{equation}
The cell state update, combining previous and new information:
\begin{equation}
    c_t = f_t \odot c_{t-1} + i_t \odot \tilde{c}_t,
\end{equation}
where $\tilde{c}_t = \tanh(W_c x_t + U_c h_{t-1} + b_c)$ represents candidate cell state values.

\subsection*{TKAN Architecture}
The TKAN architecture represents a more specialized approach, developed specifically for volume forecasting tasks. It implements a dual memory mechanism by combining Recurring Kolmogorov-Arnold Networks (RKAN) with additional memory management components. This structure enables sophisticated temporal pattern recognition through hierarchical processing.

At its core, RKAN layers process the input through layer-specific memory states. For each layer $l$, the input transformation is:
\begin{equation}
    s_{l,t}=W_{l,\tilde{x}} x_t + W_{l,\tilde{h}} \tilde{h}_{l,t-1}
\end{equation}
where $W_{l,\tilde{x}} \in \mathbb{R}^{d \times \text{KAN}_{in}}$ and $W_{l,\tilde{h}} \in \mathbb{R}^{\text{KAN}_{out} \times \text{KAN}_{in}}$ transform the current input and the previous sub-state, respectively. The RKAN layer output is computed as:

\begin{equation}
    \tilde{o}_{t} = \phi_{l}(s_{l,t}),
\end{equation}
where $\phi_l$ represents the KAN layer transformation.
The sub-layer memory state is updated according to:
\begin{equation}
    \tilde{h}_{l,t} = W_{hh} \tilde{h}_{l,t-1} + W_{hz} \tilde{o}_{t}.
\end{equation}
The full TKAN combines these RKAN outputs through:
\begin{equation}
    r_t = \text{Concat}[\phi_1(s_{1,t}),\phi_2(s_{2,t}),...,\phi_L(s_{L,t})].
\end{equation}
The output gate then processes this concatenated representation:
\begin{equation}
    o_t = \sigma(W_{o}r_t + b_o),
\end{equation}
where $W_{o} \in \mathbb{R}^{(\text{KAN}_{out} \cdot L,out)}$. The final hidden state update is given by:
\begin{equation}
    h_t = o_t \odot \tanh(c_t).
\end{equation}

This architecture implements two levels of memory management: the RKAN sub-layer memory states ($\tilde{h}_{l,t}$) capture local temporal patterns within each layer, while the global cell state ($c_t$) and hidden state ($h_t$) maintain longer-term dependencies across the entire network. This dual memory mechanism enables our model to effectively capture both fine-grained market dynamics and longer-term trading patterns, providing a rich representation for subsequent volume adjustments.

\subsection{Dynamic Volume Adjustment Network}

The core innovation of the dynamic VWAP framework is the sequential volume adjustment mechanism, which translates temporal patterns from the recurrent network into optimal execution trajectories. For each time step in the prediction horizon, a separate transformation function \(f_i\) (with \(i \in \{1, \dots, h-1\}\)) is employed, implemented as a multi-layer perceptron. For \(t > 0\), the function processes an augmented input that concatenates the RNN hidden state with the history of previous volume decisions:
\begin{equation}
    f_i(h_t, v_{1:t-1}) = W^{(3)}_i \cdot \text{ReLU}(W^{(2)}_i \cdot \text{ReLU}(W^{(1)}_i [h_t; v_{1:t-1}] + b^{(1)}_i) + b^{(2)}_i) + b^{(3)}_i,
\end{equation}
where $[h_t; v_{1:t-1}]$ represents the concatenation of the hidden state $h_t$ with the sequence of previously allocated volumes $v_{1:t-1}$. For the initial timestep ($t=0$), only the hidden state is used as input:
\begin{equation}
    f_0(h_t) = W^{(3)}_0 \cdot \text{ReLU}(W^{(2)}_0 \cdot \text{ReLU}(W^{(1)}_0 h_t + b^{(1)}_0) + b^{(2)}_0) + b^{(3)}_0.
\end{equation}
The output of these functions is then used to compute the volume adjustment factor:
\begin{equation}
    \alpha_i = 1 + \tanh(f_i(h_{l+i-1}, v_{1:i-1})).
\end{equation}
This architectural design enables each adjustment decision to be informed not only by the market conditions (captured in the hidden state) but also by the actual execution trajectory. The use of separate transformation functions for each timestep allows the model to learn time-specific responses that consider both the current market state and the accumulated execution history. This is particularly important as trading decisions often need to balance immediate market conditions with the execution progress achieved so far.

\subsection{Sequential Volume Allocation}

The volume allocation process implements a sequential mechanism that ensures both causality and volume conservation. The process can be broken down into three key steps:
\subsubsection{Base Volume Initialization}
The base volume curve $v_b$ is initialized uniformly and maintained as a learnable parameter with positivity and sum-to-one constraints:
\begin{equation}
    v_b^{(0)} = \{\frac{1}{h}, \frac{1}{h}, ..., \frac{1}{h}\}, \quad \sum_{i=1}^h v_b^{(i)} = 1, \quad v_b^{(i)} \geq 0.
\end{equation}
\subsubsection{Sequential Adjustment Process}
For timesteps $i = 1$ to $h-1$, the allocated volume is computed via a constrained adjustment of the base curve:
\begin{equation}
    v_i = \operatorname{clip}\bigl(\alpha_i\, v_b^{(i)}, 0, 1 - \sum_{j=1}^{i-1} v_j\bigr).
\end{equation}
The clip function enforces three essential constraints:
1. Non-negativity: $v_i \geq 0$
2. Running sum constraint: $\sum_{j=1}^i v_j \leq 1$
3. Volume conservation: ensures sufficient volume remains for future timesteps
Each decision incorporates both the current market state through the RNN hidden state and the execution history through the accumulated volume curve, enabling more informed and context-aware trading decisions.

\subsubsection{Final Timestep Allocation}
For the final timestep $h$, the remaining volume is allocated to ensure complete volume conservation:
\begin{equation}
    v_h = 1 - \sum_{i=1}^{h-1} v_i.
\end{equation}
This construction guarantees that the resulting volume curve satisfies all required constraints:
\begin{equation}
    \sum_{i=1}^h v_i = 1, \quad v_i \geq 0 \quad \forall i \in \{1,...,h\}
\end{equation}
This sequential adjustment mechanism has several important properties. First, it ensures that the executed volume curve always satisfies the VWAP benchmarking constraints. Second, by conditioning each adjustment on the partial volume curve, it allows the model to dynamically adapt to its own prior decisions. Finally, by preserving the base volume curve as a foundation, it strikes a balance between adaptability and robustness, mitigating the risk of extreme adjustments that could lead to poor execution quality.

\subsection{From Training to Real-Time Deployment}

A notable characteristic of the model architecture is the seamless transition from training to real-time deployment. During training, the input sequence includes complete information for the entire period, including future timesteps. This complete sequence is essential for both parameter optimization and loss function computation, as VWAP performance can only be evaluated over the full execution period.

In real-time deployment, the sequential nature of the architecture, where predictions at time $t$ depend solely on information up to that point, permits straightforward adaptation. For deployment, the unknown future portion of the input sequence is padded with zeros:

\begin{equation}
    \tilde{x}_t = \begin{cases}
        x_t & \text{if } t \leq t_{\text{current}} \\
        0 & \text{if } t > t_{\text{current}}
    \end{cases}
\end{equation}

This padding maintains the required input dimensionality while ensuring that future information does not affect current predictions. The causality of the architecture guarantees that these padded values do not influence the predictions for the current timestep, allowing the same model architecture to be used for both training and deployment without structural modifications or performance degradation.

\section{Empirical Results}

This section presents the key findings from the empirical evaluation of the dynamic VWAP framework across a range of market conditions and benchmarks. An overview of the dataset and experimental setup is provided, followed by a detailed analysis of model performance in terms of VWAP slippage, volume curve fit, and execution stability. The approach is then compared to several baseline models, and the implications of the findings for practical VWAP execution are discussed.

\subsection{Experimental Setup}

\subsubsection{Dataset and Preprocessing}

The empirical analysis utilizes hourly trading data from five major cryptocurrencies (BTC, ETH, BNB, ADA, XRP) traded on Binance perpetual futures contracts. The dataset spans from the inception of each contract until July 1, 2024, offering a comprehensive view across different market regimes and conditions. A robust data partitioning strategy is employed to ensure realistic evaluation of model performance. The final 20\% of each cryptocurrency's data is reserved for testing, providing a true out-of-sample evaluation that mirrors real-world deployment scenarios. From the remaining 80\%, the last 20\% is allocated for validation, with the remainder used for training. Due to the temporal nature of the model's lookback window, randomization in dataset splitting is avoided to prevent unwanted overlap between training and validation data.

\subsubsection{Feature Engineering}

The model processes market data using a lookback window of 120 time steps to generate predictions for the subsequent 12 periods. The feature engineering approach emphasizes stationarity and careful handling of temporal dependencies to avoid forward-looking bias. The feature set comprises:

1. Volume Features: Raw volumes are normalized by dividing by a two-week rolling average, with the averaging window shifted by the combined lookback and prediction horizon to prevent information leakage. This normalization helps capture relative volume patterns while maintaining stationarity.

2. Temporal Indicators: Cyclic patterns are incorporated through hour-of-day and day-of-week indicators, enabling the model to learn time-dependent trading patterns characteristic of cryptocurrency markets.

3. Market Metrics: The feature set includes returns computed from bin VWAP prices. Periods without trading activity are assigned zero returns, ensuring continuous representation of market state while accurately reflecting periods of market inactivity.

\subsubsection{Objective}
Given the use of hourly data, the model's objective is to predict trading proportions for each hour. Three different optimization functions are compared. The first two compute deviations from market VWAP using absolute and quadratic terms, respectively. For these calculations, the price of each bin is considered to be its VWAP, allowing for a comparison of the achieved price with the market price by weighting the bin prices using either the predicted volume curve or the market volume curve over the execution period. In addition, the optimization of the quadratic distance to the market volume curve is evaluated, enabling a comparison between direct VWAP optimization and volume curve optimization, which is more common in the literature.

\subsubsection{Model Training Configuration}

A rigorous training protocol is implemented to ensure model stability and convergence. The optimization process employs the Adam optimizer with an initial learning rate of 0.001, along with two essential callback mechanisms:

1. Early Stopping: Training terminates automatically if the validation loss shows no improvement exceeding 0.00001 over a 10-epoch window. This mechanism helps prevent overfitting while ensuring sufficient model convergence.

2. Learning Rate Adaptation: A learning rate reduction callback monitors validation loss and applies a reduction factor of 0.25 after 5 epochs without improvement. The learning rate has a lower bound of 0.000025 to maintain stable optimization.

\medskip

Training proceeds with a batch size of 128 samples and can continue up to 1000 epochs, although early stopping typically results in earlier convergence. To ensure robust evaluation of model performance and stability, 5 independent training runs are conducted for each model configuration, each with different random initializations.

\subsubsection{Implementation Details}

All models are implemented using the Keras 3 framework with a Jax backend. To ensure reproducibility, a global random seed (seed = 1) is set at the framework level. The implementation leverages TKAN package version 0.4.3 and temporal\_linear\_network version 0.1.2 (for the TLN components in the static VWAP implementation). 

\newpage

\subsection{Results and Discussion}

\begin{table}[H]
    \centering
    \caption{VWAP Optimization Results for 12 steps ahead and 120 lookback window}
    \small
    \resizebox{0.93\textwidth}{!}{%
        \begin{tabular}{llcccccccccc}
        \hline
        Model Type & Asset & Optimization & \multicolumn{2}{c}{Abs. VWAP Loss ($10^{-2}$)} & \multicolumn{2}{c}{Quad. VWAP Loss ($10^{-4}$)} & \multicolumn{2}{c}{R² Vol. Curve} & \multicolumn{2}{c}{Training Time (s)} \\
         &  & Function & Mean & Std & Mean & Std & Mean & Std & Mean & Std \\
        \hline
        Naive & BTC & N/A & 0.15874311 & 0.00000000 & 0.08780818 & 0.00000000 & 0.00000000 & 0.00000000 & 0.00000000 & 0.00000000 \\
        StaticVWAP with TLN & BTC & Absolute & 0.11927138 & 0.00061126 & 0.04950001 & 0.00093375 & -0.15073566 & 0.02958581 & 6.58394294 & 1.14440368 \\
        StaticVWAP with TLN & BTC & Quadratic & 0.12114690 & 0.00081411 & 0.04740549 & 0.00113008 & -0.36392904 & 0.05453835 & 5.71664810 & 0.08569509 \\
        StaticVWAP with TLN & BTC & Volume Curve & 0.14936921 & 0.00141546 & 0.08490292 & 0.00124393 & 0.13865099 & 0.00765448 & 7.92339087 & 0.95352642 \\
        StaticVWAP with TKAN & BTC & Absolute & 0.11684203 & 0.00049303 & 0.04770554 & 0.00110660 & -0.11334908 & 0.02263134 & 168.46392 & 12.35629 \\
        StaticVWAP with TKAN & BTC & Quadratic & 0.11801709 & 0.00106458 & 0.04544745 & 0.00191821 & -0.29856431 & 0.03826654 & 153.33461 & 0.50330536 \\
        StaticVWAP with TKAN & BTC & Volume Curve & 0.14894322 & 0.00094589 & 0.08426092 & 0.00082297 & 0.15269206 & 0.00287044 & 222.30086 & 17.19178 \\
        StaticVWAP with LSTM & BTC & Absolute & 0.11976055 & 0.00104861 & 0.05120017 & 0.00132101 & -0.10830410 & 0.02111277 & 51.63447261 & 1.38568229 \\
        StaticVWAP with LSTM & BTC & Quadratic & 0.11937241 & 0.00019484 & 0.04678259 & 0.00075100 & -0.36264214 & 0.01592363 & 50.57027316 & 0.17817769 \\
        StaticVWAP with LSTM & BTC & Volume Curve & 0.14695242 & 0.00082630 & 0.08318198 & 0.00226543 & 0.16022438 & 0.00244476 & 65.28424587 & 8.21955171 \\
        DynamicVWAP with TKAN & BTC & Absolute & \textbf{0.10525936} & 0.00050465 & \textbf{0.04129920} & 0.00158598 & -0.25283452 & 0.05233427 & 208.71017 & 18.33248 \\
        DynamicVWAP with TKAN & BTC & Quadratic & 0.11716457 & 0.00228313 & 0.04235737 & 0.00141263 & -0.43806944 & 0.12618593 & 184.70359 & 3.21572885 \\
        DynamicVWAP with TKAN & BTC & Volume Curve & 0.14388469 & 0.00459499 & 0.10851773 & 0.00543851 & 0.53748146 & 0.00295036 & 221.52546 & 15.73279 \\
        DynamicVWAP with LSTM & BTC & Absolute & 0.10575007 & 0.00127960 & 0.04707270 & 0.00240716 & -0.20447943 & 0.02173252 & 82.18055816 & 6.41036860 \\
        DynamicVWAP with LSTM & BTC & Quadratic & 0.11349924 & 0.00043251 & 0.04315868 & 0.00109093 & -0.32484988 & 0.05102079 & 71.19897299 & 2.04876808 \\
        DynamicVWAP with LSTM & BTC & Volume Curve & 0.14023268 & 0.00094018 & 0.10459246 & 0.00199253 & \textbf{0.54282326} & 0.00464932 & 84.61098056 & 6.70062535 \\
        \hline
        Naive & ETH & N/A & 0.17775816 & 0.00000000 & 0.11619556 & 0.00000000 & 0.00000000 & 0.00000000 & 0.00000000 & 0.00000000 \\
        StaticVWAP with TLN & ETH & Absolute & 0.13896361 & 0.00026868 & 0.07704772 & 0.00035661 & -0.13596994 & 0.01986146 & 6.26097364 & 0.23096691 \\
        StaticVWAP with TLN & ETH & Quadratic & 0.13992102 & 0.00023689 & 0.07361006 & 0.00095498 & -0.28744147 & 0.08706499 & 5.96797128 & 0.08868830 \\
        StaticVWAP with TLN & ETH & Volume Curve & 0.16925061 & 0.00141120 & 0.12173233 & 0.00142317 & 0.11599889 & 0.00433650 & 7.38808718 & 0.63245337 \\
        StaticVWAP with TKAN & ETH & Absolute & 0.13632348 & 0.00012737 & 0.07288946 & 0.00091962 & -0.15745594 & 0.02997837 & 175.85840 & 27.47713 \\
        StaticVWAP with TKAN & ETH & Quadratic & 0.13691810 & 0.00057300 & 0.06898466 & 0.00128563 & -0.28609347 & 0.04469012 & 149.76217 & 0.09836786 \\
        StaticVWAP with TKAN & ETH & Volume Curve & 0.16797782 & 0.00191563 & 0.11966284 & 0.00275318 & 0.12669437 & 0.00131510 & 175.65949 & 23.11001 \\
        StaticVWAP with LSTM & ETH & Absolute & 0.13908274 & 0.00066764 & 0.07739173 & 0.00251693 & -0.12567050 & 0.01243688 & 51.80121160 & 5.49519323 \\
        StaticVWAP with LSTM & ETH & Quadratic & 0.13874996 & 0.00071484 & 0.07273126 & 0.00260816 & -0.29236349 & 0.04784417 & 50.22743292 & 1.95464302 \\
        StaticVWAP with LSTM & ETH & Volume Curve & 0.16757467 & 0.00206867 & 0.12106892 & 0.00281386 & 0.13295122 & 0.00231076 & 65.96278644 & 6.58184341 \\
        DynamicVWAP with TKAN & ETH & Absolute & \textbf{0.12090351} & 0.00105061 &  \textbf{0.05820394} & 0.00606626 & -0.36164006 & 0.07777891 & 218.17821 & 31.41944 \\
        DynamicVWAP with TKAN & ETH & Quadratic & 0.13420704 & 0.00048353 & 0.06155421 & 0.00376874 & -0.34871240 & 0.06998671 & 179.72871 & 2.26166123 \\
        DynamicVWAP with TKAN & ETH & Volume Curve & 0.15213849 & 0.00895828 & 0.14477868 & 0.01503425 & 0.55199230 & 0.00349011 & 206.74531 & 9.29955299 \\
        DynamicVWAP with LSTM & ETH & Absolute & 0.12130527 & 0.00203506 & 0.06049004 & 0.00346033 & -0.20793096 & 0.05407269 & 72.49890113 & 4.89879187 \\
        DynamicVWAP with LSTM & ETH & Quadratic & 0.13092618 & 0.00079659 & 0.06327784 & 0.00224571 & -0.24169885 & 0.05874037 & 68.90146255 & 2.57381396 \\
        DynamicVWAP with LSTM & ETH & Volume Curve & 0.15238284 & 0.00329357 & 0.14955543 & 0.00706529 & \textbf{0.55604589} & 0.00665185 & 85.64025993 & 11.04057245 \\
        \hline
        Naive & BNB & N/A & 0.17411649 & 0.00000000 & 0.11887422 & 0.00000000 & 0.00000000 & 0.00000000 & 0.00000000 & 0.00000000 \\
        StaticVWAP with TLN & BNB & Absolute & 0.13769057 & 0.00056687 & 0.07674849 & 0.00114790 & -0.16782647 & 0.02039261 & 5.97753563 & 0.40601649 \\
        StaticVWAP with TLN & BNB & Quadratic & 0.14058632 & 0.00108101 & 0.07105021 & 0.00041829 & -0.41619455 & 0.04845878 & 5.82552795 & 0.10969033 \\
        StaticVWAP with TLN & BNB & Volume Curve & 0.16627431 & 0.00097884 & 0.11731241 & 0.00124564 & 0.08507098 & 0.00348360 & 5.88534303 & 0.31191838 \\
        StaticVWAP with TKAN & BNB & Absolute & 0.13779818 & 0.00068007 & 0.07304682 & 0.00208107 & -0.20178974 & 0.04904818 & 152.08773 & 7.79358244 \\
        StaticVWAP with TKAN & BNB & Quadratic & 0.14064121 & 0.00111225 & 0.06984355 & 0.00124068 & -0.40046372 & 0.05459933 & 143.01995 & 0.22680267 \\
        StaticVWAP with TKAN & BNB & Volume Curve & 0.16389677 & 0.00104582 & 0.11523929 & 0.00112501 & 0.09884318 & 0.00225406 & 168.13328 & 18.25175085 \\
        StaticVWAP with LSTM & BNB & Absolute & 0.13728492 & 0.00044532 & 0.07574741 & 0.00173882 & -0.16223255 & 0.03287896 & 47.47297196 & 0.96557658 \\
        StaticVWAP with LSTM & BNB & Quadratic & 0.14207184 & 0.00095548 & 0.06938407 & 0.00100383 & -0.42748443 & 0.03219933 & 46.91346102 & 0.13752507 \\
        StaticVWAP with LSTM & BNB & Volume Curve & 0.16160807 & 0.00163531 & 0.11258962 & 0.00176982 & 0.10063113 & 0.00237855 & 53.57504334 & 4.27191039 \\
        DynamicVWAP with TKAN & BNB & Absolute & \textbf{0.12928014} & 0.00097290 & \textbf{0.06106497} & 0.00481985 & -0.41639036 & 0.12995663 & 212.81138 & 19.68545669 \\
        DynamicVWAP with TKAN & BNB & Quadratic & 0.13816078 & 0.00108385 & 0.06230573 & 0.00116104 & -0.43768488 & 0.03573618 & 171.20077 & 2.42834300 \\
        DynamicVWAP with TKAN & BNB & Volume Curve & 0.15831196 & 0.00237909 & 0.13827127 & 0.00401578 & 0.48004346 & 0.00763932 & 188.22576 & 10.09825448 \\
        DynamicVWAP with LSTM & BNB & Absolute & 0.13075633 & 0.00116716 & 0.06752896 & 0.00198895 & -0.34261975 & 0.09547927 & 78.22121482 & 5.21327253 \\
        DynamicVWAP with LSTM & BNB & Quadratic & 0.13613380 & 0.00134886 & 0.05984929 & 0.00091537 & -0.40282933 & 0.04356278 & 66.18081899 & 2.08406464 \\
        DynamicVWAP with LSTM & BNB & Volume Curve & 0.15895155 & 0.00096154 & 0.14315926 & 0.00148856 & \textbf{0.48645234} & 0.00480968 & 71.81402125 & 3.62312253 \\
        \hline
        Naive & ADA & N/A & 0.22620659 & 0.00000000 & 0.22880934 & 0.00000000 & 0.00000000 & 0.00000000 & 0.00000000 & 0.00000000 \\
        StaticVWAP with TLN & ADA & Absolute & 0.18802823 & 0.00053372 & 0.15834919 & 0.00592556 & -0.19308613 & 0.05353767 & 6.08805647 & 0.33938803 \\
        StaticVWAP with TLN & ADA & Quadratic & 0.19592619 & 0.00085794 & 0.13822354 & 0.00134807 & -0.55848189 & 0.06839654 & 5.88964295 & 0.15442150 \\
        StaticVWAP with TLN & ADA & Volume Curve & 0.22112048 & 0.00287925 & 0.24471748 & 0.00527829 & 0.08645346 & 0.00221476 & 6.03375468 & 0.57368560 \\
        StaticVWAP with TKAN & ADA & Absolute & 0.18606430 & 0.00041329 & 0.14990953 & 0.00279260 & -0.20398289 & 0.02184775 & 160.72590 & 10.16107 \\
        StaticVWAP with TKAN & ADA & Quadratic & 0.19504803 & 0.00215119 & 0.13538916 & 0.00240247 & -0.47462739 & 0.06013042 & 144.76179 & 2.17708601 \\
        StaticVWAP with TKAN & ADA & Volume Curve & 0.21688870 & 0.00168358 & 0.24487790 & 0.00179522 & 0.10853112 & 0.00247803 & 161.22746 & 23.32622 \\
        StaticVWAP with LSTM & ADA & Absolute & 0.18714230 & 0.00104794 & 0.15833926 & 0.00567292 & -0.15773954 & 0.02168280 & 48.94532022 & 1.88230979 \\
        StaticVWAP with LSTM & ADA & Quadratic & 0.19649315 & 0.00177664 & 0.13473901 & 0.00208179 & -0.53246783 & 0.05352562 & 47.29803710 & 0.11642053 \\
        StaticVWAP with LSTM & ADA & Volume Curve & 0.21915285 & 0.00101610 & 0.24978559 & 0.00316859 & 0.10998640 & 0.00335903 & 57.04135704 & 1.27692198 \\
        DynamicVWAP with TKAN & ADA & Absolute & 0.17767891 & 0.00272218 & 0.12740992 & 0.00930824 & -0.45691523 & 0.14637307 & 177.85913 & 7.30870941 \\
        DynamicVWAP with TKAN & ADA & Quadratic & 0.19642825 & 0.00400710 & \textbf{0.12433632} & 0.00453353 & -0.68882630 & 0.12299890 & 171.59860 & 2.61905030 \\
        DynamicVWAP with TKAN & ADA & Volume Curve & 0.21115856 & 0.00535849 & 0.31867027 & 0.01747973 & 0.50884105 & 0.00370986 & 211.98463 & 13.05221 \\
        DynamicVWAP with LSTM & ADA & Absolute & \textbf{0.17594782} & 0.00086925 & 0.13341191 & 0.00504261 & -0.37247308 & 0.06933388 & 67.87838511 & 2.51005713 \\
        DynamicVWAP with LSTM & ADA & Quadratic & 0.19386384 & 0.00403683 & 0.13384703 & 0.00223607 & -0.53517202 & 0.09480017 & 65.50151954 & 0.40602272 \\
        DynamicVWAP with LSTM & ADA & Volume Curve & 0.21547738 & 0.00263136 & 0.33619204 & 0.00512109 & \textbf{0.51746946} & 0.00377054 & 83.79096918 & 6.92280483 \\
        \hline
        Naive & XRP & N/A & 0.22385468 & 0.00000000 & 0.27592453 & 0.00000000 & 0.00000000 & 0.00000000 & 0.00000000 & 0.00000000 \\
        StaticVWAP with TLN & XRP & Absolute & 0.18250506 & 0.00075665 & 0.18804847 & 0.00134658 & -0.26329018 & 0.01229442 & 5.89494438 & 0.17577209 \\
        StaticVWAP with TLN & XRP & Quadratic & 0.18838942 & 0.00134069 & 0.16050903 & 0.00451912 & -0.70089282 & 0.12680138 & 5.81431966 & 0.08779065 \\
        StaticVWAP with TLN & XRP & Volume Curve & 0.21978152 & 0.00111825 & 0.29401255 & 0.00307838 & 0.05323325 & 0.00298841 & 6.20426478 & 0.58831923 \\
        StaticVWAP with TKAN & XRP & Absolute & 0.18163190 & 0.00030957 & 0.18449907 & 0.00290836 & -0.25769478 & 0.02345099 & 187.10690 & 10.47068 \\
        StaticVWAP with TKAN & XRP & Quadratic & 0.18653250 & 0.00055580 & 0.15111423 & 0.00222610 & -0.61209932 & 0.02288775 & 146.89900 & 2.17488309 \\
        StaticVWAP with TKAN & XRP & Volume Curve & 0.21654904 & 0.00253169 & 0.29347703 & 0.00556594 & 0.07188528 & 0.00112462 & 179.31275 & 19.89412 \\
        StaticVWAP with LSTM & XRP & Absolute & 0.18236685 & 0.00051911 & 0.18671843 & 0.00337798 & -0.25814381 & 0.01314251 & 49.65237107 & 1.28548545 \\
        StaticVWAP with LSTM & XRP & Quadratic & 0.18699214 & 0.00115349 & 0.15528060 & 0.00330169 & -0.66035077 & 0.06968230 & 47.95475988 & 0.09837191 \\
        StaticVWAP with LSTM & XRP & Volume Curve & 0.21691068 & 0.00099450 & 0.29703385 & 0.00249483 & 0.07276691 & 0.00237884 & 55.17223105 & 3.60249138 \\
        DynamicVWAP with TKAN & XRP & Absolute & \textbf{0.17419565} & 0.00138844 & 0.17708713 & 0.00558218 & -0.39368663 & 0.04917958 & 217.44334 & 17.52560 \\
        DynamicVWAP with TKAN & XRP & Quadratic & 0.18598679 & 0.00192495 & \textbf{0.14618580} & 0.00389745 & -0.69780867 & 0.09215657 & 174.79140 & 2.40635431 \\
        DynamicVWAP with TKAN & XRP & Volume Curve & 0.20582119 & 0.00327031 & 0.36081353 & 0.01217642 & 0.47710991 & 0.00500007 & 198.12152 & 11.01618 \\
        DynamicVWAP with LSTM & XRP & Absolute & 0.17437634 & 0.00063044 & 0.18289676 & 0.00545318 & -0.25355316 & 0.03504690 & 68.67374697 & 2.01133037 \\
        DynamicVWAP with LSTM & XRP & Quadratic & 0.18645420 & 0.00088455 & 0.15269616 & 0.00266152 & -0.74541301 & 0.06363387 & 67.83486104 & 2.40201454 \\
        DynamicVWAP with LSTM & XRP & Volume Curve & 0.21010165 & 0.00359678 & 0.38042180 & 0.00823294 & \textbf{0.49037368} & 0.00266708 & 72.44338455 & 3.48800860 \\
        \hline
        \end{tabular}
    }
    \label{tab:dynamic_vwap_results}
\end{table}
\clearpage

The empirical analysis in table \ref{tab:dynamic_vwap_results} reveals significant improvements in VWAP execution performance through dynamic modeling approaches. The models are evaluated using three key metrics: absolute VWAP loss (measured in $10^{-2}$), quadratic VWAP loss (measured in $10^{-4}$), and the \(R^2\) score for volume curve prediction. The \(R^2\) score is noteworthy as it represents the traditional optimization target for volume-centric methodologies and serves as an important benchmark for comparing the approach with conventional methods.

\subsubsection{Performance Across Model Architectures}

The analysis shows a clear progression of performance improvements as the models evolve from simple linear models to more sophisticated architectures. The original static VWAP implementation using Temporal Linear Networks (TLN) demonstrates significant improvements over naive baselines, with reductions in absolute VWAP loss ranging from 20–25\% across all assets.

\medskip

The transition from TLN to recurrent architectures in the static framework yields substantial additional improvements. Looking at ETH, the static LSTM reduces absolute VWAP loss to 0.139 compared to TLN's 0.144, representing a 3.5\% improvement. The TKAN architecture further enhances these results, with the static implementation achieving 0.136, a 5.6\% improvement over TLN. Similar patterns emerge for BNB, where static LSTM and TKAN achieve losses of 0.137 and 0.134 respectively, compared to TLN's 0.142, representing improvements of 3.5\% and 5.6\%. This consistent pattern suggests that recurrent architectures' ability to capture temporal dependencies provides meaningful advantages even in static implementations. The transition to dynamic architectures marks an even more significant advancement in performance. For ETH, the DynamicVWAP with TKAN achieves an absolute VWAP loss of 0.121, an 11.0\% improvement over its static counterpart (0.136) and a 16.0\% improvement over the static TLN baseline (0.144). In BNB markets, the dynamic TKAN implementation reduces absolute VWAP loss to 0.129, representing improvements of 3.7\% over static TKAN (0.134) and 9.2\% over static TLN (0.142). The dynamic LSTM shows similar patterns but consistently falls slightly behind TKAN, with losses of 0.123 for ETH and 0.131 for BNB. The superiority of TKAN over LSTM becomes particularly pronounced in the dynamic setting. For ADA, dynamic TKAN achieves an absolute loss of 0.178 compared to dynamic LSTM's 0.184, a 3.3\% improvement. This advantage extends to quadratic loss, where dynamic TKAN's 0.127 represents a 4.8\% improvement over dynamic LSTM's 0.133. XRP shows similar patterns, with dynamic TKAN outperforming dynamic LSTM by 3.1\% in absolute loss (0.174 versus 0.179) and 3.2\% in quadratic loss (0.177 versus 0.183).

\medskip

These improvements are particularly noteworthy as they scale with market liquidity. In highly liquid markets such as BTC and ETH, the transition from static to dynamic architectures yields the largest relative improvements, often exceeding 10\%. For less liquid assets such as ADA and XRP, while the absolute improvements are smaller, the relative advantage of dynamic over static implementations remains consistent, indicating that the approach successfully adapts to varying market conditions. The enhanced performance of TKAN in both static and dynamic settings, particularly its superior ability to capture complex temporal patterns in real time, establishes a clear hierarchy of performance. Dynamic TKAN consistently leads, followed by dynamic LSTM, then static TKAN, static LSTM, and finally static TLN, with this ordering remaining stable across all assets and market conditions.

\subsubsection{Impact of Optimization Functions}

In many optimization scenarios, quadratic loss is favored because its gradient decreases with the magnitude of the error, providing smoother parameter updates and facilitating convergence. In contrast, absolute loss maintains a constant gradient regardless of the error size, which may lead to instability in the updates as the same magnitude of adjustment is applied for both small and large errors. Surprisingly, experimental results indicate that models trained with absolute VWAP loss consistently outperform those trained with quadratic loss in out-of-sample testing, for both static and dynamic implementations. For example, in BTC experiments, dynamic TKAN models exhibited a 10.2\% improvement in absolute VWAP loss when trained with absolute loss rather than quadratic loss. Moreover, models trained with absolute loss not only achieved lower absolute VWAP loss but also produced better performance when evaluated using quadratic metrics.

\medskip

Dynamic models optimized for volume curve prediction achieved remarkable improvements in \(R^2\) scores compared to their static counterparts. For instance, with ETH, the dynamic LSTM attained an \(R^2\) of 0.56, compared to 0.13 for the static version—an improvement ranging from 200\% to 400\% across different assets. Despite this enhanced predictive power for the volume curve, such improvement does not directly translate into superior VWAP execution performance. In fact, models that directly optimize for the VWAP execution target, particularly when using absolute loss, yield the best overall execution results—even if their \(R^2\) scores for volume prediction are lower or negative (for example, an \(R^2\) of -0.25 for BTC in the dynamic TKAN model).

\medskip

It is possible that quadratic loss, by emphasizing larger errors, may overweight extreme events during training, leading to an excessive drift in the internal base volume curve and reduced robustness under market stress. In contrast, the constant gradient provided by absolute loss appears to offer a more stable training regime, which is reflected in better generalization to out-of-sample data. In summary, while quadratic loss typically offers smoother optimization dynamics, the robustness achieved by directly targeting absolute execution error appears to be more beneficial in the context of VWAP execution.

\subsubsection{Asset-Specific Performance}

Market structure and liquidity significantly influence the relative advantages of dynamic over static models. More liquid assets like BTC and ETH show the largest relative improvements when moving from static to dynamic approaches, with the performance enhancement being particularly stable across different market conditions. For less liquid assets like ADA and XRP, while the absolute performance remains weaker than for major assets, the relative improvement from static to dynamic models remains consistent, suggesting the proposed approach successfully adapts to varying market conditions.

\subsubsection{Computational Considerations}

The computational requirements of the different architectures warrant careful consideration, particularly with respect to their performance benefits. The transition from static to dynamic models introduces a modest increase in computational overhead, primarily because both approaches already incorporate sequential processing of the full lookback window through recurrent neural network components. For TKAN implementations, dynamic models require 150–220 seconds of training time compared to 115–170 seconds for static counterparts, representing an approximate 30\% increase. Similarly, dynamic LSTM models require 50–85 seconds compared to 35–65 seconds for static versions.

\medskip
A more significant computational distinction is observed when comparing recurrent architectures with the simple linear approach of TLN, which exhibits training times of only 5–8 seconds. This order-of-magnitude difference in computational requirements between TLN and recurrent implementations (both static and dynamic) represents the primary computational trade-off. However, the increased computational cost is justified by substantial performance benefits, as recurrent architectures reduce VWAP tracking errors by 15–25\% compared to TLN approaches.

\medskip

These computational considerations should be evaluated in the context of practical VWAP execution, where even marginal improvements in execution quality can result in significant financial benefits given typical transaction volumes in cryptocurrency markets. Furthermore, the observed training times remain within practical limits for regular model updates, even accounting for the increased complexity of recurrent architectures. The moderate additional overhead introduced by dynamic implementations over static ones appears justified given the resulting improvements in execution quality.

\subsubsection{Visual Analysis of Execution Performance}
\begin{figure}[H]
    \centering
    \includegraphics[width=\columnwidth]{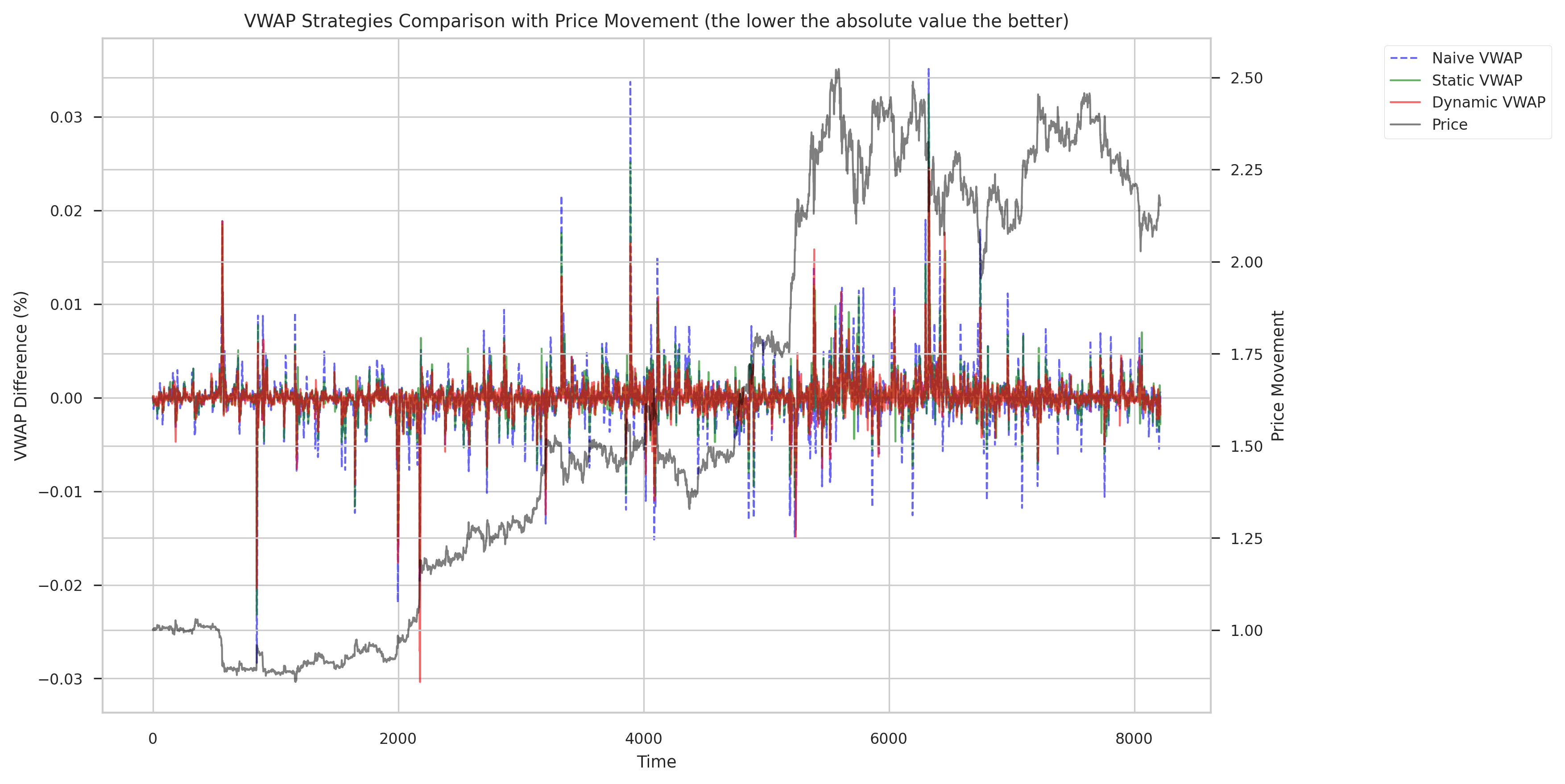}
    \caption{Slippage between approaches on the full out-of-sample set}
    \label{fig:dynamic_slippage_full}
\end{figure}

\begin{figure}[H]
    \centering
    \includegraphics[width=\columnwidth]{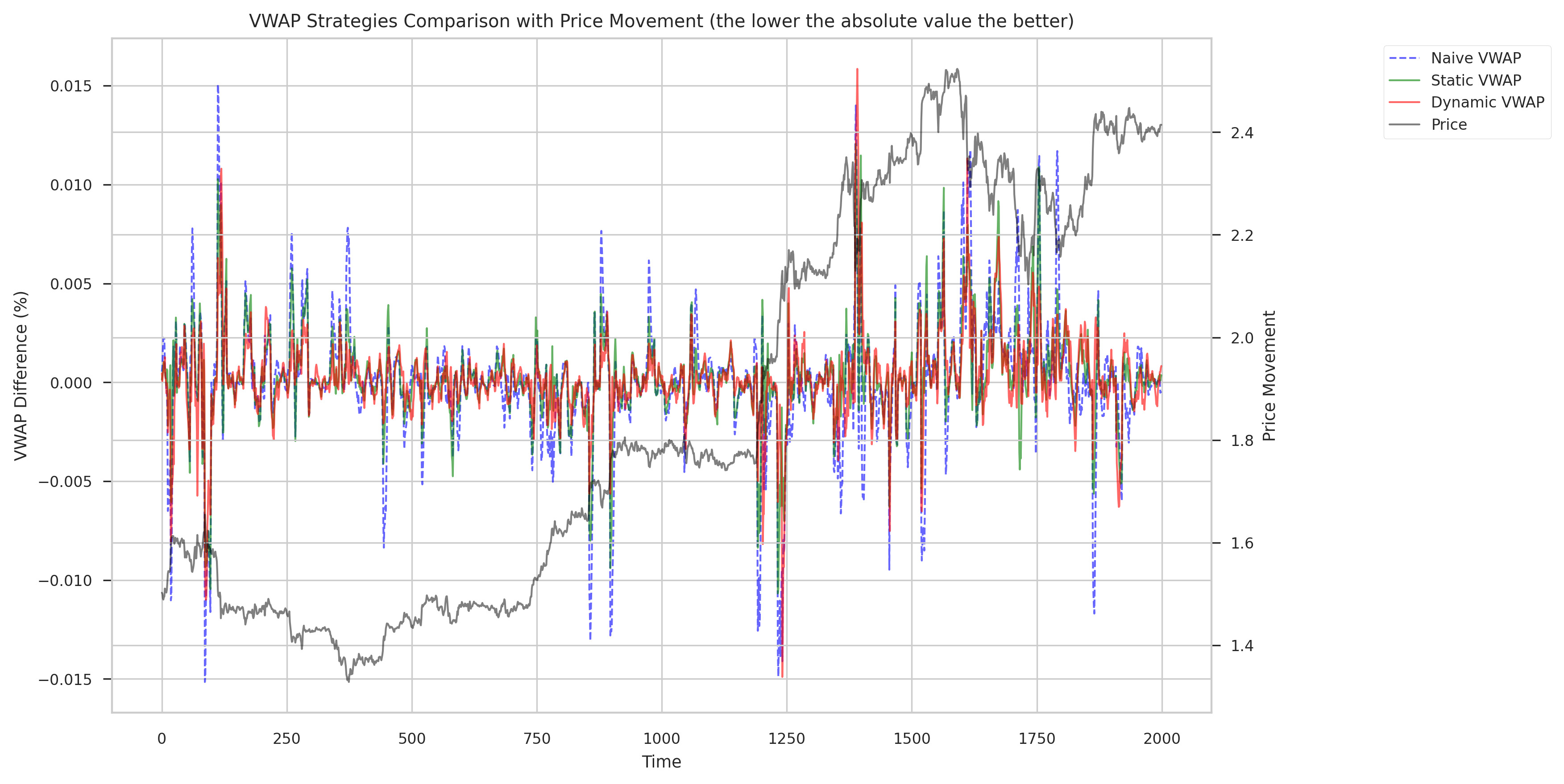}
    \caption{Slippage between approaches on a subsample of the out-of-sample set}
    \label{fig:dynamic_slippage_subset}
\end{figure}

\begin{figure}[H]
    \centering
    \includegraphics[width=\columnwidth]{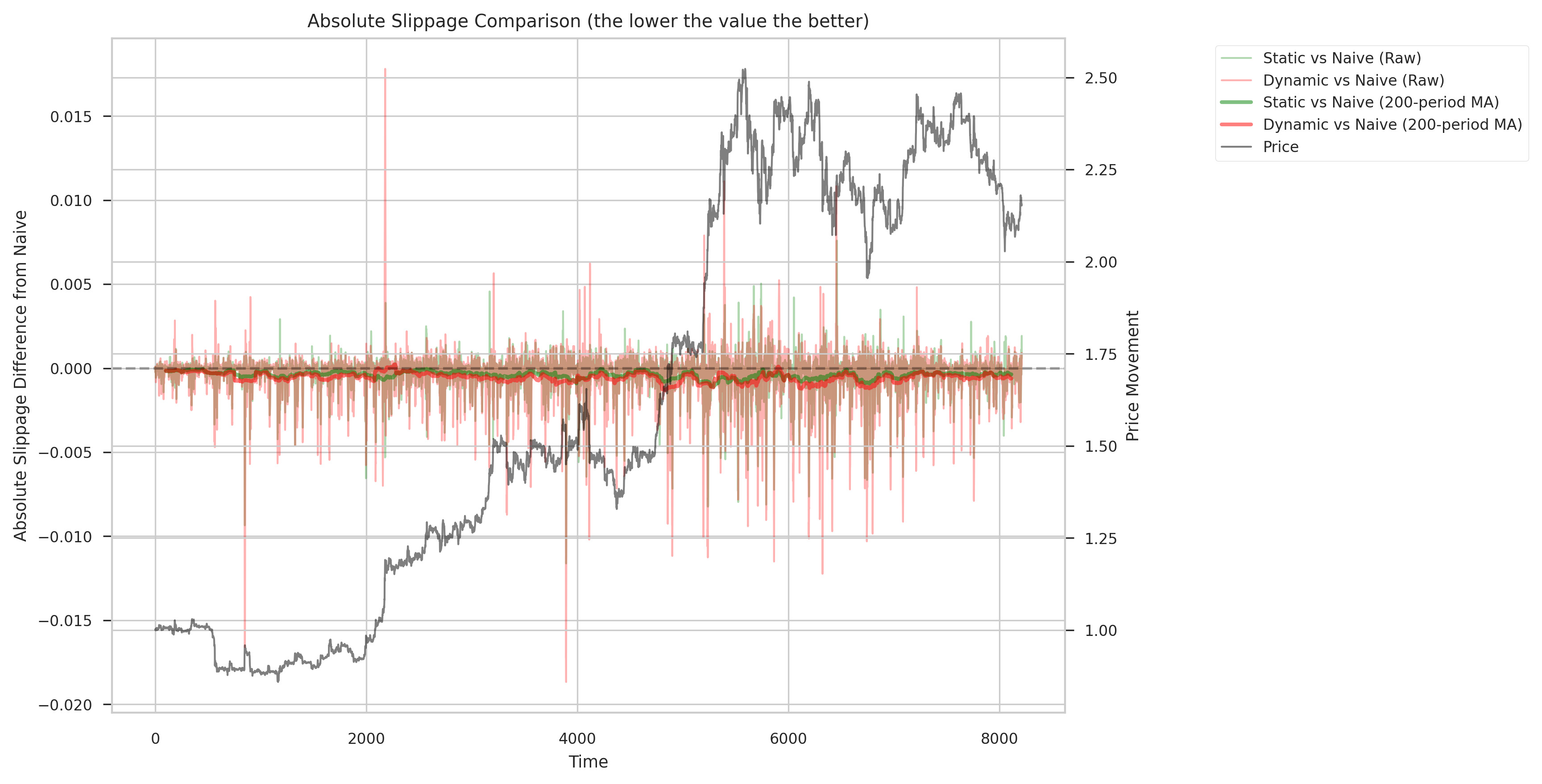}
    \caption{Difference in absolute slippage versus naive approach on the full out-of-sample set. Negative values indicate improved performance over the naive approach.}
    \label{fig:dynamic_slippage_diff_full}
\end{figure}

\begin{figure}[H]
    \centering
    \includegraphics[width=\columnwidth]{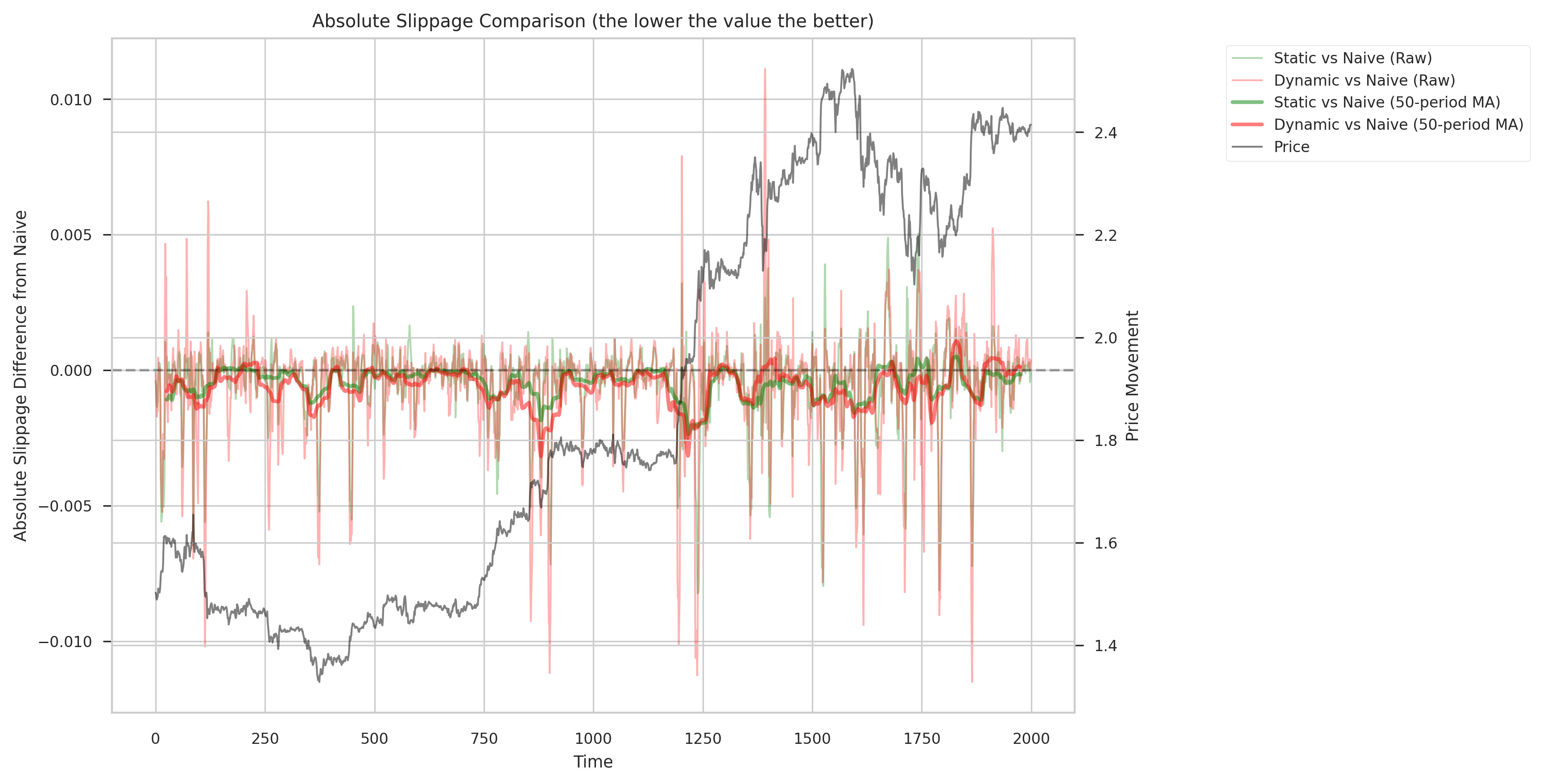}
    \caption{Difference in absolute slippage versus naive approach on the subsample set. Negative values indicate improved performance over the naive approach.}
    \label{fig:dynamic_slippage_diff_subset}
\end{figure}

Graphical analysis provides additional insights into the comparative performance of different VWAP execution approaches. Figures \ref{fig:dynamic_slippage_full} and \ref{fig:dynamic_slippage_subset} present a comparison of slippage between naive, static, and dynamic implementations across different time horizons. Figures \ref{fig:dynamic_slippage_diff_full} and \ref{fig:dynamic_slippage_diff_subset} highlight the relative performance improvements over the naive baseline.

\medskip

In Figure \ref{fig:dynamic_slippage_full}, examining the full sample period, a notable pattern emerges during periods of sharp price movements. The magnitude of slippage spikes demonstrates a clear hierarchy of performance: dynamic implementations consistently exhibit smaller deviations compared to static models, which in turn show reduced slippage relative to the naive approach. This pattern is particularly evident during significant market events, suggesting that more sophisticated models better maintain execution quality during challenging conditions.

However, the density of information in the full sample analysis can obscure performance differences during more stable market periods. To address this, Figure \ref{fig:dynamic_slippage_subset} provides a more granular view of execution performance. This analysis reveals a consistent pattern in which dynamic implementations not only achieve lower absolute slippage but also maintain this advantage consistently. The visualization clearly demonstrates that improvements in slippage reduction from dynamic models (compared to naive) consistently exceed those achieved by static implementations, providing visual confirmation of the quantitative metrics presented in the earlier tables.

To address potential obscuring of performance differences during more stable market periods, Figure \ref{fig:dynamic_slippage_subset} provides a more granular view of execution performance. This analysis reveals a consistent pattern in which dynamic implementations not only achieve lower absolute slippage but also maintain this advantage consistently. The visualization confirms that improvements in slippage reduction from dynamic models exceed those achieved by static implementations, aligning with the quantitative metrics presented earlier.

Figures \ref{fig:dynamic_slippage_diff_full} and \ref{fig:dynamic_slippage_diff_subset} quantify these improvements by showing the difference in absolute slippage versus the naive approach. The predominantly negative values for dynamic implementations (shown in red and blue) confirm systematic outperformance, with the magnitude of improvement often exceeding 0.005 (50 basis points) during challenging market conditions.

\subsection{Analysis of Extended Time Horizons}
The appendix Tables \ref{table:dynamic_48_step} and \ref{table:dynamic_6_step} provide insights into model performance across different execution horizons, comparing 48-step and 6-step ahead predictions with the baseline 12-step model. Several key patterns emerge from this analysis. The short-horizon results presented in Table \ref{table:dynamic_6_step} demonstrate notably improved performance across all metrics compared to longer horizons, with absolute VWAP losses typically 30-40\% lower than in the 48-step scenario shown in Table \ref{table:dynamic_48_step}. This suggests that shorter execution windows benefit more significantly from dynamic adaptation. The relative advantage of dynamic over static implementations remains consistent across time horizons, though the absolute magnitude of improvement decreases with longer horizons. This pattern indicates that while dynamic benefits persist, the inherent difficulty of longer-term prediction partially offsets the advantages of adaptive execution. Particularly noteworthy is the stability of model performance, as evidenced by the standard deviations reported in both tables. These remain proportionally consistent across horizons, suggesting that the dynamic implementations maintain reliable execution quality regardless of the prediction timeframe. This stability is particularly important for practical applications where execution horizons may vary based on order characteristics.

\subsubsection{Visual Analysis of Execution Performance}

The graphical analysis reveals several key patterns in the execution behavior across different time horizons and model architectures. For shorter prediction windows (6 and 12 steps ahead), static VWAP implementations using recurrent neural networks demonstrate notable variability around their mean execution curves. This flexibility in predictions appears to contribute to their improved performance compared to traditional approaches. This variability becomes even more pronounced in dynamic implementations, particularly in models optimized for volume curve prediction. The enhanced adaptability manifests as wider prediction bands around the mean execution trajectory, suggesting these models can better respond to changing market conditions while maintaining overall execution objectives.
A striking observation across all architectures is the consistency of average execution curves (shown in red). Despite the significant differences in model complexity and implementation approach, these curves maintain remarkably similar patterns. This convergence suggests that the dynamic approaches preserve the fundamental principles of optimal VWAP execution while enabling better adaptation to market conditions. The analysis of longer-horizon predictions (48 steps ahead) reveals an interesting constraint effect in dynamic models. Significant contraction of prediction bands around the mean can be observed, a direct consequence of the model formulation. The use of scaling factors bounded between 0 and 2 effectively limits the variability of predictions at longer horizons. This observation suggests that while the current model specification performs well for shorter time horizons, modifications may be necessary for optimal long-horizon performance. Specifically, the constraint mechanism might benefit from horizon-dependent scaling to maintain appropriate levels of adaptability across different execution timeframes. These visual patterns provide important insights into both the strengths of the dynamic approach and potential areas for future refinement, particularly in the handling of longer execution horizons. The consistency of mean execution patterns across architectures, combined with the enhanced adaptability of dynamic models at shorter horizons, supports the quantitative findings regarding the advantages of dynamic implementation while highlighting specific areas for potential improvement in model design.

\section{Conclusion}

This paper demonstrates that dynamic neural approaches represent a significant advancement in VWAP execution methodology, particularly in cryptocurrency markets characterized by high volatility and evolving microstructure. While my framework achieves substantial improvements without explicitly modeling market impact, its modular design allows for straightforward integration of impact considerations through modification of the loss function, opening interesting avenues for future research. The empirical results reveal several key insights about the role of recurrent neural architectures in execution problems. While the initial application of RNNs to static VWAP models showed only modest improvements over traditional approaches (3-5\% reduction in tracking error), their true potential emerges through dynamic implementation. By leveraging RNNs' inherent ability to maintain and update state information, the proposed dynamic framework achieves substantial performance gains, reducing absolute VWAP tracking error by 10-15\% in liquid markets and maintaining consistent outperformance across varying market conditions. This dramatic improvement in performance when moving from static to dynamic implementations suggests that the key to successful RNN application lies not in their raw predictive power, but in their ability to adapt to evolving market conditions. The computational considerations reinforce this insight about effective architectural choices. When applied directly, RNNs introduce substantial computational overhead compared to simple linear models like TLN - training times increase by an order of magnitude, from 5-8 seconds to over 150 seconds for TKAN implementations. However, these architectures prove essential as building blocks for our dynamic framework. The additional 30\% computational overhead introduced by the dynamic implementation represents a reasonable trade-off given the resulting 10-15\% improvement in execution quality. This demonstrates that the key to achieving favorable performance-to-cost ratios lies in designing frameworks that effectively leverage the unique capabilities of sophisticated neural architectures.

\medskip

The success of this dynamic framework opens several promising directions for future research. The integration of market impact modeling through modified loss functions could provide more realistic execution strategies for large orders. Additionally, extending the model to incorporate cross-asset dependencies and exploring alternative neural architectures could further enhance performance. Significant improvements could be achieved through more sophisticated feature engineering, moving beyond our intentionally basic approach used to demonstrate the framework's potential. Development of horizon-dependent scaling factors could also enhance the model's effectiveness across different execution timeframes. These results suggest that adaptive neural architectures can effectively address the challenges of modern VWAP execution while maintaining computational efficiency suitable for practical deployment. The framework's ability to balance execution quality with computational constraints makes it particularly valuable for institutional investors managing large orders in cryptocurrency markets. By implementing our model as a standard Keras architecture and making all code publicly available, I provide a foundation that practitioners can readily deploy and researchers can build upon. This research thus contributes to both the theoretical understanding of optimal execution and practical implementation of VWAP strategies, bridging the gap between academic research and industry practice in an increasingly important area of market microstructure.

\section*{Code Availability}
The source code used for all experiments and analyses in this paper is available at \url{https://github.com/remigenet/DeepDynamicVWAP}.

\newpage
\bibliographystyle{IEEEtran}
\bibliography{bib}

\begin{thebibliography}{10}
\providecommand{\url}[1]{#1}
\csname url@samestyle\endcsname
\providecommand{\newblock}{\relax}
\providecommand{\bibinfo}[2]{#2}
\providecommand{\BIBentrySTDinterwordspacing}{\spaceskip=0pt\relax}
\providecommand{\BIBentryALTinterwordstretchfactor}{4}
\providecommand{\BIBentryALTinterwordspacing}{\spaceskip=\fontdimen2\font plus
\BIBentryALTinterwordstretchfactor\fontdimen3\font minus \fontdimen4\font\relax}
\providecommand{\BIBforeignlanguage}[2]{{%
\expandafter\ifx\csname l@#1\endcsname\relax
\typeout{** WARNING: IEEEtran.bst: No hyphenation pattern has been}%
\typeout{** loaded for the language `#1'. Using the pattern for}%
\typeout{** the default language instead.}%
\else
\language=\csname l@#1\endcsname
\fi
#2}}
\providecommand{\BIBdecl}{\relax}
\BIBdecl

\bibitem{genet2025staticvwap}
R.~Genet, ``Deep learning for vwap execution in crypto markets: Beyond the volume curve,'' \emph{arXiv preprint arXiv:2502.13722}, 2025.

\bibitem{genet2024tln}
R.~Genet and H.~Inzirillo, ``A temporal linear network for time series forecasting,'' 2024, arXiv preprint arXiv:2410.21448.

\bibitem{Mackenzie}
M.~Mackenzie, ``High frequency trading under scrutiny,'' \emph{Financial Times}, no.~1, july 2009.

\bibitem{perold1988implementation}
A.~F. Perold, ``The implementation shortfall: Paper versus reality,'' \emph{Journal of Portfolio Management}, vol.~14, no.~3, p.~4, 1988.

\bibitem{TotalCostOfTransactions}
S.~Berkowitz, D.~Logue, and E.~Noser, ``The total cost of transactions on the nyse,'' \emph{Journal of Finance}, vol.~43, pp. 97--112, 1988.

\bibitem{Konishi}
\BIBentryALTinterwordspacing
H.~Konishi, ``Optimal slice of a vwap trade,'' \emph{Journal of Financial Markets}, vol.~5, no.~2, pp. 197--221, 2002. [Online]. Available: \url{https://EconPapers.repec.org/RePEc:eee:finmar:v:5:y:2002:i:2:p:197-221}
\BIBentrySTDinterwordspacing

\bibitem{Culoch2007}
\BIBentryALTinterwordspacing
J.~McCulloch and V.~Kazakov, ``Optimal vwap trading strategy and relative volume,'' no. 201, 2007. [Online]. Available: \url{https://EconPapers.repec.org/RePEc:uts:rpaper:201}
\BIBentrySTDinterwordspacing

\bibitem{Humphery}
M.~L. Humphery-Jenner, ``{Optimal VWAP trading under noisy conditions},'' \emph{Journal of Banking \& Finance}, vol.~35, no.~9, pp. 2319--2329, September 2011.

\bibitem{LeFol2006}
\BIBentryALTinterwordspacing
J.~Bialkowski, S.~Darolles, and G.~Le~Fol, ``{Improving VWAP strategies: A dynamic volume approach},'' \emph{{Journal of Banking and Finance}}, vol.~32, pp. 1709--1722, 2008. [Online]. Available: \url{https://halshs.archives-ouvertes.fr/halshs-00676946}
\BIBentrySTDinterwordspacing

\bibitem{frei}
C.~Frei and N.~Westray, ``Optimal execution of a vwap order: A stochastic control approach,'' \emph{Mathematical Finance}, vol.~25, 10 2013.

\bibitem{Madhavan2002}
A.~Madhavan, ``Vwap strategies, transaction performance: The changing face of trading investment guides series,'' \emph{Institutional Investor Inc.}, pp. 32--38, 2002.

\bibitem{Easley1987}
\BIBentryALTinterwordspacing
D.~Easley and M.~O'Hara, ``Price, trade size, and information in securities markets,'' \emph{Journal of Financial Economics}, vol.~19, no.~1, pp. 69--90, 1987. [Online]. Available: \url{https://EconPapers.repec.org/RePEc:eee:jfinec:v:19:y:1987:i:1:p:69-90}
\BIBentrySTDinterwordspacing

\bibitem{Foster}
S.~Viswanathan and F.~Foster, ``A theory of the interday variations in volume, variance, and trading costs in securities markets,'' \emph{Review of Financial Studies}, vol.~3, pp. 593--624, 02 1990.

\bibitem{Tauchen1983}
\BIBentryALTinterwordspacing
G.~Tauchen and M.~Pitts, ``The price variability-volume relationship on speculative markets,'' \emph{Econometrica}, vol.~51, no.~2, pp. 485--505, 1983. [Online]. Available: \url{https://EconPapers.repec.org/RePEc:ecm:emetrp:v:51:y:1983:i:2:p:485-505}
\BIBentrySTDinterwordspacing

\bibitem{Karpoff1987}
J.~Karpoff, ``The relation between price changes and trading volume: A survey,'' \emph{Journal of Financial and Quantitative Analysis}, vol.~22, no.~1, pp. 109--126, 1987.

\bibitem{Gourieroux}
C.~Gourieroux, J.~Jasiak, and G.~Le~Fol, ``{Intra-day market activity},'' \emph{Journal of Financial Markets}, vol.~2, no.~3, pp. 193--226, August 1999.

\bibitem{Culoch2012}
\BIBentryALTinterwordspacing
J.~McCulloch and V.~Kazakov, ``Mean variance optimal vwap trading,'' April 2012. [Online]. Available: \url{https://ssrn.com/abstract=1803858}
\BIBentrySTDinterwordspacing

\bibitem{LeFol2012}
\BIBentryALTinterwordspacing
J.~Bialkowski, S.~Darolles, and G.~Le~Fol, ``{Reducing the risk of VWAP orders execution - A new approach to modeling intra-day volume},'' \emph{{JASSA}}, no.~1, 2012. [Online]. Available: \url{https://hal.archives-ouvertes.fr/hal-01632822}
\BIBentrySTDinterwordspacing

\bibitem{bouchard}
\BIBentryALTinterwordspacing
B.~Bouchard and N.~M. Dang, ``{Generalized stochastic target problems for pricing and partial hedging under loss constraints - Application in optimal book liquidation},'' \emph{{Finance and Stochastics}}, vol.~17, no.~1, pp. 31--72, Jan. 2013. [Online]. Available: \url{https://hal.archives-ouvertes.fr/hal-00508291}
\BIBentrySTDinterwordspacing

\bibitem{Tianhui}
\BIBentryALTinterwordspacing
R.~Carmona and T.~Li, ``Dynamic programming and trade execution,'' 2013. [Online]. Available: \url{http://arks.princeton.edu/ark:/88435/dsp01w0892b00k}
\BIBentrySTDinterwordspacing

\bibitem{Gueant}
O.~Gu{\'e}ant and G.~Royer, ``Vwap execution and guaranteed vwap,'' \emph{SIAM J. Financial Math.}, vol.~5, pp. 445--471, 2014.

\bibitem{sezer2020financial}
O.~B. Sezer, M.~U. Gudelek, and A.~M. Ozbayoglu, ``Financial time series forecasting with deep learning: A systematic literature review: 2005--2019,'' \emph{Applied Soft Computing}, vol.~90, p. 106181, 2020.

\bibitem{hochreiter1997}
S.~Hochreiter and J.~Schmidhuber, ``Long short-term memory,'' \emph{Neural computation}, vol.~9, no.~8, pp. 1735--1780, 1997.

\bibitem{cho2014}
K.~Cho, B.~Van~Merri{\"e}nboer, C.~Gulcehre, D.~Bahdanau, F.~Bougares, H.~Schwenk, and Y.~Bengio, ``Learning phrase representations using rnn encoder-decoder for statistical machine translation,'' in \emph{Proceedings of the 2014 Conference on Empirical Methods in Natural Language Processing (EMNLP)}, 2014, pp. 1724--1734.

\bibitem{bahdanau2014neural}
D.~Bahdanau, K.~Cho, and Y.~Bengio, ``Neural machine translation by jointly learning to align and translate,'' \emph{arXiv preprint arXiv:1409.0473}, 2014.

\bibitem{vaswani2017attention}
A.~Vaswani, N.~Shazeer, N.~Parmar, J.~Uszkoreit, L.~Jones, A.~N. Gomez, {\L}.~Kaiser, and I.~Polosukhin, ``Attention is all you need,'' \emph{Advances in neural information processing systems}, vol.~30, 2017.

\bibitem{ackerer2020deep}
D.~Ackerer, N.~Tagasovska, and T.~Vatter, ``Deep smoothing of the implied volatility surface,'' in \emph{34th Conference on Neural Information Processing Systems (NeurIPS 2020)}, Vancouver, Canada, 2020.

\bibitem{horvath2019deep}
B.~Horvath, A.~Muguruza, and M.~Tomas, ``Deep learning volatility: A deep neural network perspective on pricing and calibration in (rough) volatility models,'' \emph{arXiv preprint arXiv:1901.09647}, 2019.

\bibitem{zhang2023deep}
C.~Zhang, N.~N.~A. Sjarif, and R.~Ibrahim, ``Deep learning models for price forecasting of financial time series: A review of recent advancements: 2020-2022,'' \emph{arXiv preprint arXiv:2305.04811}, 2023.

\bibitem{genet2024tkan}
R.~Genet and H.~Inzirillo, ``Tkan: Temporal kolmogorov-arnold networks,'' \emph{arXiv preprint arXiv:2405.07344}, 2024.

\bibitem{inzirillo2024sigkan}
H.~Inzirillo and R.~Genet, ``Sigkan: Signature-weighted kolmogorov-arnold networks for time series,'' \emph{arXiv preprint arXiv:2406.17890}, 2024.

\bibitem{genet2024tkat}
R.~Genet and H.~Inzirillo, ``A temporal kolmogorov-arnold transformer for time series forecasting,'' \emph{ArXiv}, 2024.

\bibitem{inzirillo2024kamoe}
H.~Inzirillo and R.~Genet, ``A gated residual kolmogorov-arnold networks for mixtures of experts,'' 2024, arXiv preprint arXiv:2409.15161.

\bibitem{genet2025siggate}
R.~Genet and H.~Inzirillo, ``Siggate: Enhancing recurrent neural networks with signature-based gating mechanisms,'' \emph{arXiv preprint arXiv:2502.09318}, 2025.

\bibitem{papanicolaou2023optimal}
A.~Papanicolaou, H.~Fu, P.~Krishnamurthy, B.~Healy, and F.~Khorrami, ``An optimal control strategy for execution of large stock orders using lstms,'' \emph{arXiv preprint arXiv:2301.09705}, 2023.

\end{thebibliography}

\newpage
\begin{appendix}
\section{appendix}
\subsection{48 steps table results}
\begin{table}[H]
    \centering
    \caption{VWAP Optimization Results for 48 steps ahead and 120 lookback window}
    \small
    \resizebox{0.93\textwidth}{!}{%
        \begin{tabular}{llcccccccccc}
        \hline
        Model Type & Asset & Optimization & \multicolumn{2}{c}{Abs. VWAP Loss ($10^{-2}$)} & \multicolumn{2}{c}{Quad. VWAP Loss ($10^{-4}$)} & \multicolumn{2}{c}{R² Vol. Curve} & \multicolumn{2}{c}{Training Time (s)} \\
         &  & Function & Mean & Std & Mean & Std & Mean & Std & Mean & Std \\
        \hline
        Naive & BTC & N/A & 0.29455480 & 0.00000000 & 0.24016084 & 0.00000000 & 0.00000000 & 0.00000000 & 0.00000000 & 0.00000000 \\
        StaticVWAP with TLN & BTC & Absolute & 0.20996968 & 0.00396063 & 0.11913272 & 0.00586642 & -0.13468582 & 0.06504267 & 7.19253902 & 0.87950195 \\
        StaticVWAP with TLN & BTC & Quadratic & 0.21867290 & 0.00324281 & 0.12195075 & 0.00277514 & -0.38380868 & 0.14105786 & 6.53614035 & 0.18242597 \\
        StaticVWAP with TLN & BTC & Volume Curve & 0.26934916 & 0.00450855 & 0.20882210 & 0.00579848 & 0.15792550 & 0.00339095 & 6.12318420 & 0.19666206 \\
        StaticVWAP with TKAN & BTC & Absolute & 0.20947179 & 0.00108181 & 0.11802824 & 0.00200633 & -0.06826087 & 0.00838763 & 186.15744 & 16.62807 \\
        StaticVWAP with TKAN & BTC & Quadratic & 0.21669798 & 0.00085301 & 0.11883552 & 0.00126366 & -0.25756119 & 0.04437503 & 156.99467 & 1.78377 \\
        StaticVWAP with TKAN & BTC & Volume Curve & 0.26410604 & 0.00463460 & 0.20244789 & 0.00630356 & 0.17110288 & 0.00520260 & 154.67395 & 0.95658 \\
        StaticVWAP with LSTM & BTC & Absolute & 0.22008915 & 0.00403345 & 0.13848941 & 0.00498093 & -0.10647021 & 0.03002864 & 53.31942 & 2.15936 \\
        StaticVWAP with LSTM & BTC & Quadratic & 0.21982396 & 0.00284685 & 0.12414330 & 0.00554233 & -0.28508129 & 0.02491924 & 51.50950 & 0.09892 \\
        StaticVWAP with LSTM & BTC & Volume Curve & 0.26497862 & 0.00664891 & 0.20575495 & 0.01094346 & 0.17857434 & 0.00385339 & 51.03363 & 0.17281 \\
        DynamicVWAP with TKAN & BTC & Absolute & 0.20344035 & 0.00545412 & 0.11518467 & 0.01450806 & -0.08005180 & 0.10187624 & 349.83393 & 16.68234 \\
        DynamicVWAP with TKAN & BTC & Quadratic & 0.21243012 & 0.00241512 & \textbf{0.10948813} & 0.00398090 & -0.23538589 & 0.07632112 & 343.67843 & 6.41085 \\
        DynamicVWAP with TKAN & BTC & Volume Curve & 0.25673180 & 0.00910208 & 0.24049124 & 0.01602785 & \textbf{0.49357890} & 0.00563921 & 358.71979 & 15.61475 \\
        DynamicVWAP with LSTM & BTC & Absolute & 0.20568359 & 0.00683018 & 0.15062606 & 0.01512616 & -0.26539066 & 0.04471402 & 210.22449 & 10.10632 \\
        DynamicVWAP with LSTM & BTC & Quadratic & \textbf{0.20065229} & 0.00303235 & 0.11093245 & 0.00558061 & -0.22355913 & 0.07174325 & 201.23169 & 5.87284 \\
        DynamicVWAP with LSTM & BTC & Volume Curve & 0.25743768 & 0.01161459 & 0.24885090 & 0.02242506 & 0.49238944 & 0.00412532 & 204.10822 & 3.33619 \\
        \hline
        Naive & ETH & N/A & 0.30678847 & 0.00000000 & 0.29848686 & 0.00000000 & 0.00000000 & 0.00000000 & 0.00000000 & 0.00000000 \\
        StaticVWAP with TLN & ETH & Absolute & 0.23257165 & 0.00154950 & 0.18060520 & 0.00326324 & -0.08442341 & 0.03231453 & 6.59268355 & 0.63376119 \\
        StaticVWAP with TLN & ETH & Quadratic & 0.24165091 & 0.00433838 & 0.18290560 & 0.00873870 & -0.22880010 & 0.08464580 & 6.34411855 & 0.08198686 \\
        StaticVWAP with TLN & ETH & Volume Curve & 0.28871828 & 0.00597324 & 0.28942457 & 0.01059617 & 0.13453966 & 0.00167982 & 6.06883922 & 0.09956466 \\
        StaticVWAP with TKAN & ETH & Absolute & 0.23239572 & 0.00301913 & 0.18060783 & 0.00546645 & -0.13837383 & 0.00495348 & 182.33204 & 24.47089 \\
        StaticVWAP with TKAN & ETH & Quadratic & 0.24680996 & 0.00181747 & 0.18169032 & 0.00533692 & -0.27267007 & 0.02707372 & 152.46992 & 1.89479 \\
        StaticVWAP with TKAN & ETH & Volume Curve & 0.27747250 & 0.00753112 & 0.26976960 & 0.01434693 & 0.13998114 & 0.00191305 & 150.44388 & 0.22603 \\
        StaticVWAP with LSTM & ETH & Absolute & 0.24109511 & 0.00473620 & 0.19237855 & 0.00837183 & -0.15332445 & 0.01613196 & 51.80458 & 1.56369 \\
        StaticVWAP with LSTM & ETH & Quadratic & 0.24113376 & 0.00327896 & 0.17180547 & 0.00529873 & -0.28844500 & 0.07102014 & 50.74356 & 0.37879 \\
        StaticVWAP with LSTM & ETH & Volume Curve & 0.27665633 & 0.00453206 & 0.27387518 & 0.00709642 & 0.14796637 & 0.00309746 & 50.29155 & 0.13305 \\
        DynamicVWAP with TKAN & ETH & Absolute & \textbf{0.22744370} & 0.00751542 & 0.17162873 & 0.01857800 & -0.18619999 & 0.05344997 & 350.77260 & 14.14173 \\
        DynamicVWAP with TKAN & ETH & Quadratic & 0.23394556 & 0.00221355 & \textbf{0.14776996} & 0.00203905 & -0.26662744 & 0.03358505 & 335.93586 & 4.42816 \\
        DynamicVWAP with TKAN & ETH & Volume Curve & 0.27511278 & 0.01057361 & 0.32987165 & 0.02355857 & 0.51679585 & 0.00504112 & 398.45288 & 49.43910 \\
        DynamicVWAP with LSTM & ETH & Absolute & 0.23551905 & 0.00515884 & 0.20000990 & 0.01168122 & -0.22472650 & 0.06890893 & 204.81041 & 4.69640 \\
        DynamicVWAP with LSTM & ETH & Quadratic & 0.22799444 & 0.00319051 & 0.16352156 & 0.01000850 & -0.32725163 & 0.01812173 & 197.41560 & 1.21897 \\
        DynamicVWAP with LSTM & ETH & Volume Curve & 0.26579678 & 0.01638573 & 0.31281018 & 0.03545075 & \textbf{0.51727627} & 0.00940737 & 212.50978 & 15.18331 \\
        \hline
        Naive & BNB & N/A & 0.30650598 & 0.00000000 & 0.27640975 & 0.00000000 & 0.00000000 & 0.00000000 & 0.00000000 & 0.00000000 \\
        StaticVWAP with TLN & BNB & Absolute & 0.23793343 & 0.00150448 & 0.15793566 & 0.00081481 & -0.39200412 & 0.25155706 & 6.42121563 & 0.45187214 \\
        StaticVWAP with TLN & BNB & Quadratic & 0.27409320 & 0.00338686 & 0.19695878 & 0.00557209 & -0.89363862 & 0.20849643 & 6.30463729 & 0.19243534 \\
        StaticVWAP with TLN & BNB & Volume Curve & 0.28011040 & 0.00338951 & 0.24163632 & 0.00580465 & 0.09265859 & 0.00439849 & 5.71810126 & 0.13697845 \\
        StaticVWAP with TKAN & BNB & Absolute & 0.23746534 & 0.00210910 & 0.15866478 & 0.00130691 & -0.16483657 & 0.00458007 & 153.75189 & 8.81126 \\
        StaticVWAP with TKAN & BNB & Quadratic & 0.27597650 & 0.00382458 & 0.19434714 & 0.00367297 & -0.38958528 & 0.03019198 & 146.28291 & 1.85219 \\
        StaticVWAP with TKAN & BNB & Volume Curve & 0.27928966 & 0.00377729 & 0.24434605 & 0.00703057 & 0.09622988 & 0.00157368 & 145.34921 & 1.89071 \\
        StaticVWAP with LSTM & BNB & Absolute & 0.24657206 & 0.00220006 & 0.17253058 & 0.00534732 & -0.20835604 & 0.01937689 & 50.23494 & 1.33559 \\
        StaticVWAP with LSTM & BNB & Quadratic & 0.27237946 & 0.00727144 & 0.19601306 & 0.00831399 & -0.38557689 & 0.04695624 & 48.83598 & 0.12263 \\
        StaticVWAP with LSTM & BNB & Volume Curve & 0.27900550 & 0.00341841 & 0.24639693 & 0.00606209 & 0.10276128 & 0.00159999 & 48.45848 & 0.14040 \\
        DynamicVWAP with TKAN & BNB & Absolute & \textbf{0.22824177} & 0.00271431 & \textbf{0.13578534} & 0.00283858 & -0.37803197 & 0.14007530 & 364.12167 & 40.53192 \\
        DynamicVWAP with TKAN & BNB & Quadratic & 0.25862965 & 0.00153851 & 0.15485217 & 0.00516124 & -0.46916227 & 0.02471390 & 328.39294 & 4.32234 \\
        DynamicVWAP with TKAN & BNB & Volume Curve & 0.27361990 & 0.00722314 & 0.26158658 & 0.01626400 & 0.44207561 & 0.00889953 & 331.99762 & 5.37810 \\
        DynamicVWAP with LSTM & BNB & Absolute & 0.23677337 & 0.00670602 & 0.16743426 & 0.01604389 & -0.32430545 & 0.09489654 & 202.39413 & 14.40988 \\
        DynamicVWAP with LSTM & BNB & Quadratic & 0.23624948 & 0.00329866 & 0.13817477 & 0.00354395 & -0.43337581 & 0.04527596 & 193.77293 & 2.36557 \\
        DynamicVWAP with LSTM & BNB & Volume Curve & 0.28905886 & 0.01175479 & 0.30667922 & 0.02917550 & \textbf{0.44973522} & 0.00465319 & 196.12166 & 5.96231 \\
        \hline
        Naive & ADA & N/A & 0.41902370 & 0.00000000 & 0.66928870 & 0.00000000 & 0.00000000 & 0.00000000 & 0.00000000 & 0.00000000 \\
        StaticVWAP with TLN & ADA & Absolute & 0.35045750 & 0.00103442 & 0.43756717 & 0.00170681 & -0.26117657 & 0.04346317 & 6.40609212 & 0.46196553 \\
        StaticVWAP with TLN & ADA & Quadratic & 0.37224630 & 0.00228324 & 0.40961605 & 0.00199164 & -0.59824312 & 0.21583582 & 6.15162325 & 0.10177388 \\
        StaticVWAP with TLN & ADA & Volume Curve & 0.40926003 & 0.00172651 & 0.73642105 & 0.00600985 & 0.08303814 & 0.00266709 & 5.75634294 & 0.12379043 \\
        StaticVWAP with TKAN & ADA & Absolute & 0.34884828 & 0.00065148 & 0.42447720 & 0.00695541 & -0.25535890 & 0.01085621 & 156.78983 & 8.19544 \\
        StaticVWAP with TKAN & ADA & Quadratic & 0.37249547 & 0.00424976 & 0.40915403 & 0.00500919 & -0.46060049 & 0.05197456 & 146.92128 & 2.22436 \\
        StaticVWAP with TKAN & ADA & Volume Curve & 0.40681228 & 0.00755595 & 0.72950774 & 0.02134474 & 0.09405450 & 0.00139185 & 146.21703 & 1.98754 \\
        StaticVWAP with LSTM & ADA & Absolute & 0.36142403 & 0.00381785 & 0.46792775 & 0.00636657 & -0.29728518 & 0.02417225 & 49.86250 & 1.19051 \\
        StaticVWAP with LSTM & ADA & Quadratic & 0.38161750 & 0.01058896 & 0.44141635 & 0.00513752 & -0.56532033 & 0.07377473 & 48.50581 & 0.11233 \\
        StaticVWAP with LSTM & ADA & Volume Curve & 0.40199107 & 0.00454153 & 0.73438350 & 0.01841021 & 0.09204840 & 0.00130554 & 48.32261 & 0.09353 \\
        DynamicVWAP with TKAN & ADA & Absolute & \textbf{0.33821172} & 0.00403377 & 0.41263090 & 0.02717904 & -0.60698202 & 0.05401519 & 398.48977 & 24.45133 \\
        DynamicVWAP with TKAN & ADA & Quadratic & 0.36987490 & 0.00600796 & \textbf{0.36169320} & 0.00286300 & -0.41057710 & 0.05149440 & 335.88603 & 1.68949 \\
        DynamicVWAP with TKAN & ADA & Volume Curve & 0.39972162 & 0.01727822 & 0.86128790 & 0.05011240 & 0.47853491 & 0.00575267 & 355.76760 & 25.34387 \\
        DynamicVWAP with LSTM & ADA & Absolute & 0.3606668 & 0.00086925 & 0.57424027 & 0.00504261 & -0.37247308 & 0.06933388 & 67.87839 & 2.51006 \\
        DynamicVWAP with LSTM & ADA & Quadratic & 0.38473535 & 0.00612871 & 0.50609810 & 0.04279679 & -0.59027868 & 0.04959509 & 200.49852 & 5.01619 \\
        DynamicVWAP with LSTM & ADA & Volume Curve & 0.40072864 & 0.00654358 & 0.92620766 & 0.02400405 & \textbf{0.49023477} & 0.00612646 & 200.49965 & 4.10699 \\
        \hline
        Naive & XRP & N/A & 0.41742540 & 0.00000000 & 0.65299230 & 0.00000000 & 0.00000000 & 0.00000000 & 0.00000000 & 0.00000000 \\
        StaticVWAP with TLN & XRP & Absolute & 0.33705962 & 0.00081158 & 0.39610392 & 0.00486802 & -0.29163120 & 0.06865979 & 6.88782787 & 0.27537009 \\
        StaticVWAP with TLN & XRP & Quadratic & 0.36356530 & 0.00669528 & 0.34909240 & 0.01663762 & -0.93975203 & 0.19090502 & 6.45987515 & 0.09955837 \\
        StaticVWAP with TLN & XRP & Volume Curve & 0.40340160 & 0.00394753 & 0.66079790 & 0.00673557 & 0.06135918 & 0.00633603 & 5.78328662 & 0.08604623 \\
        StaticVWAP with TKAN & XRP & Absolute & 0.33504078 & 0.00397613 & 0.40053720 & 0.02645089 & -0.28508377 & 0.03846750 & 235.27480 & 44.34086 \\
        StaticVWAP with TKAN & XRP & Quadratic & 0.35795587 & 0.00308409 & 0.31800407 & 0.00187512 & -0.59862939 & 0.02014337 & 151.21649 & 3.64032 \\
        StaticVWAP with TKAN & XRP & Volume Curve & 0.39931685 & 0.00791025 & 0.65552104 & 0.02109294 & 0.06999691 & 0.00197563 & 148.46736 & 2.31226 \\
        StaticVWAP with LSTM & XRP & Absolute & 0.33964166 & 0.00595049 & 0.38865677 & 0.01578740 & -0.31425556 & 0.05313223 & 52.36923 & 2.60869 \\
        StaticVWAP with LSTM & XRP & Quadratic & 0.35538730 & 0.00372743 & 0.34142160 & 0.02874227 & -0.69351881 & 0.15188871 & 66.88039 & 16.72271 \\
        StaticVWAP with LSTM & XRP & Volume Curve & 0.39982203 & 0.00197379 & 0.67166820 & 0.00614194 & 0.07391420 & 0.00256875 & 49.30747 & 0.18741 \\
        DynamicVWAP with TKAN & XRP & Absolute & \textbf{0.33318818} & 0.00995495 & 0.41706810 & 0.05299714 & -0.36516364 & 0.06205204 & 424.50836 & 49.96236 \\
        DynamicVWAP with TKAN & XRP & Quadratic & 0.34535490 & 0.00874594 & \textbf{0.30571142} & 0.00845760 & -0.61051196 & 0.05460565 & 354.23790 & 17.21853 \\
        DynamicVWAP with TKAN & XRP & Volume Curve & 0.40637960 & 0.01033883 & 0.76814330 & 0.04227697 & 0.40304078 & 0.01093896 & 350.02209 & 9.80104 \\
        DynamicVWAP with LSTM & XRP & Absolute & 0.34658557 & 0.01235865 & 0.50474360 & 0.06043496 & -0.47486589 & 0.04153346 & 221.02721 & 23.45177 \\
        DynamicVWAP with LSTM & XRP & Quadratic & 0.33354154 & 0.00692625 & 0.30999085 & 0.01441384 & -0.68894252 & 0.13531818 & 202.74952 & 14.34388 \\
        DynamicVWAP with LSTM & XRP & Volume Curve & 0.39457900 & 0.00476943 & 0.74837583 & 0.01709443 & \textbf{0.42024922} & 0.00395634 & 206.06428 & 6.79596 \\
        \hline
        \end{tabular}
    }
    \label{table:dynamic_48_step}
\end{table}
\newpage
\subsection{6 steps table results}
\begin{table}[H]
    \centering
    \caption{VWAP Optimization Results for 6 steps ahead and 15 lookback window}
    \small
    \resizebox{0.93\textwidth}{!}{%
        \begin{tabular}{llcccccccccc}
        \hline
        Model Type & Asset & Optimization & \multicolumn{2}{c}{Abs. VWAP Loss ($10^{-2}$)} & \multicolumn{2}{c}{Quad. VWAP Loss ($10^{-4}$)} & \multicolumn{2}{c}{R² Vol. Curve} & \multicolumn{2}{c}{Training Time (s)} \\
         &  & Function & Mean & Std & Mean & Std & Mean & Std & Mean & Std \\
        \hline
        Naive & BTC & N/A & 0.10536299 & 0.00000000 & 0.04656231 & 0.00000000 & 0.00000000 & 0.00000000 & 0.00000000 & 0.00000000 \\
        StaticVWAP with TLN & BTC & Absolute & 0.08798713 & 0.00031209 & 0.03085725 & 0.00059162 & -0.16602963 & 0.01698641 & 5.70932260 & 0.58667578 \\
        StaticVWAP with TLN & BTC & Quadratic & 0.08887652 & 0.00018162 & 0.02757752 & 0.00033888 & -0.52214096 & 0.04592633 & 5.43053064 & 0.21433314 \\
        StaticVWAP with TLN & BTC & Volume Curve & 0.10309176 & 0.00066908 & 0.04851772 & 0.00066159 & 0.08195296 & 0.00081065 & 6.53815751 & 0.56167831 \\
        StaticVWAP with TKAN & BTC & Absolute & 0.08654067 & 0.00055559 & 0.02984685 & 0.00090313 & -0.21231952 & 0.05848570 & 35.11852660 & 3.59169689 \\
        StaticVWAP with TKAN & BTC & Quadratic & 0.08805765 & 0.00037871 & 0.02774281 & 0.00076116 & -0.54034200 & 0.08627341 & 31.01990805 & 0.49687609 \\
        StaticVWAP with TKAN & BTC & Volume Curve & 0.10126430 & 0.00079148 & 0.04682230 & 0.00079749 & 0.10241954 & 0.00259091 & 39.47357264 & 2.97604917 \\
        StaticVWAP with LSTM & BTC & Absolute & 0.08645161 & 0.00039586 & 0.03042696 & 0.00051130 & -0.18243823 & 0.02492621 & 11.13044186 & 0.65431721 \\
        StaticVWAP with LSTM & BTC & Quadratic & 0.08871464 & 0.00106948 & 0.02718048 & 0.00064181 & -0.66743500 & 0.14485578 & 10.50564947 & 0.12637020 \\
        StaticVWAP with LSTM & BTC & Volume Curve & 0.10130497 & 0.00040596 & 0.04697227 & 0.00048692 & 0.10591496 & 0.00121268 & 13.52486930 & 1.50308165 \\
        DynamicVWAP with TKAN & BTC & Absolute & 0.07624506 & 0.00036701 & 0.02391857 & 0.00048204 & -0.33208177 & 0.07045228 & 43.25347543 & 1.30328037 \\
        DynamicVWAP with TKAN & BTC & Quadratic & 0.08410655 & 0.00196021 & \textbf{0.02384726} & 0.00088352 & -0.64795871 & 0.07919718 & 43.90831299 & 2.21874436 \\
        DynamicVWAP with TKAN & BTC & Volume Curve & 0.09803374 & 0.00209867 & 0.05920818 & 0.00226267 & 0.43522378 & 0.00423703 & 52.96247697 & 2.65963907 \\
        DynamicVWAP with LSTM & BTC & Absolute & \textbf{0.07536691} & 0.00058447 & 0.02409948 & 0.00077079 & -0.36991840 & 0.07091915 & 19.33353643 & 2.41144563 \\
        DynamicVWAP with LSTM & BTC & Quadratic & 0.08455877 & 0.00096317 & 0.02534151 & 0.00076324 & -0.54419480 & 0.09577007 & 17.36622095 & 0.32617380 \\
        DynamicVWAP with LSTM & BTC & Volume Curve & 0.09873200 & 0.00138782 & 0.05920202 & 0.00141068 & \textbf{0.44815243} & 0.00679720 & 20.38888893 & 1.00731634 \\
        \hline
        Naive & ETH & N/A & 0.12115163 & 0.00000000 & 0.06169037 & 0.00000000 & 0.00000000 & 0.00000000 & 0.00000000 & 0.00000000 \\
        StaticVWAP with TLN & ETH & Absolute & 0.10032286 & 0.00027192 & 0.04170017 & 0.00029383 & -0.19411217 & 0.00921049 & 5.43342438 & 0.23512108 \\
        StaticVWAP with TLN & ETH & Quadratic & 0.10086425 & 0.00024519 & 0.03888644 & 0.00022257 & -0.40397870 & 0.02320226 & 5.26746883 & 0.05218601 \\
        StaticVWAP with TLN & ETH & Volume Curve & 0.11780038 & 0.00065017 & 0.06531054 & 0.00067357 & 0.07355373 & 0.00017511 & 5.95999279 & 0.60165022 \\
        StaticVWAP with TKAN & ETH & Absolute & 0.09832314 & 0.00061235 & 0.04149345 & 0.00034803 & -0.19506517 & 0.01175127 & 33.68728828 & 4.08641751 \\
        StaticVWAP with TKAN & ETH & Quadratic & 0.10020934 & 0.00075081 & 0.03745977 & 0.00066542 & -0.50555204 & 0.10062819 & 29.78190584 & 0.41293999 \\
        StaticVWAP with TKAN & ETH & Volume Curve & 0.11602328 & 0.00036207 & 0.06477320 & 0.00052188 & 0.09327520 & 0.00224688 & 34.77609081 & 3.37071872 \\
        StaticVWAP with LSTM & ETH & Absolute & 0.09811972 & 0.00014415 & 0.04204797 & 0.00019596 & -0.17637206 & 0.01639027 & 10.93825479 & 0.42286903 \\
        StaticVWAP with LSTM & ETH & Quadratic & 0.09988533 & 0.00017954 & 0.03866927 & 0.00062276 & -0.42637281 & 0.04744849 & 10.16382036 & 0.11706390 \\
        StaticVWAP with LSTM & ETH & Volume Curve & 0.11568806 & 0.00102975 & 0.06484018 & 0.00104825 & 0.09355015 & 0.00091458 & 12.51770787 & 1.77043083 \\
        DynamicVWAP with TKAN & ETH & Absolute & 0.08547299 & 0.00052161 & 0.03321413 & 0.00072963 & -0.33568026 & 0.10461838 & 46.32587733 & 4.17462042 \\
        DynamicVWAP with TKAN & ETH & Quadratic & 0.09626032 & 0.00199162 & 0.03325069 & 0.00206275 & -0.48308619 & 0.10323971 & 41.67789125 & 2.19734579 \\
        DynamicVWAP with TKAN & ETH & Volume Curve & 0.11122487 & 0.00186474 & 0.08231308 & 0.00160759 & 0.45949600 & 0.00543203 & 52.53450165 & 4.38007529 \\
        DynamicVWAP with LSTM & ETH & Absolute & \textbf{0.08522300} & 0.00078212 & \textbf{0.03183053} & 0.00098655 & -0.40996997 & 0.04346825 & 17.60930257 & 0.67233507 \\
        DynamicVWAP with LSTM & ETH & Quadratic & 0.09514316 & 0.00033152 & 0.03575233 & 0.00060645 & -0.41975853 & 0.03710844 & 16.97923551 & 0.16329749 \\
        DynamicVWAP with LSTM & ETH & Volume Curve & 0.11005618 & 0.00209386 & 0.08162374 & 0.00298013 & \textbf{0.46800165} & 0.00454018 & 23.17767649 & 0.97107500 \\
        \hline
        Naive & BNB & N/A & 0.11474912 & 0.00000000 & 0.05820391 & 0.00000000 & 0.00000000 & 0.00000000 & 0.00000000 & 0.00000000 \\
        StaticVWAP with TLN & BNB & Absolute & 0.09845396 & 0.00019751 & 0.03998742 & 0.00084720 & -0.17957728 & 0.04373623 & 6.26492701 & 2.30066363 \\
        StaticVWAP with TLN & BNB & Quadratic & 0.10214207 & 0.00057717 & 0.03688421 & 0.00027868 & -0.55171598 & 0.06011258 & 5.13236227 & 0.06179787 \\
        StaticVWAP with TLN & BNB & Volume Curve & 0.11094530 & 0.00049496 & 0.05917190 & 0.00045685 & 0.05733062 & 0.00103973 & 6.47753806 & 0.49499146 \\
        StaticVWAP with TKAN & BNB & Absolute & 0.09579776 & 0.00068800 & 0.03937370 & 0.00113767 & -0.16288673 & 0.03358899 & 32.56861877 & 3.90016350 \\
        StaticVWAP with TKAN & BNB & Quadratic & 0.10144097 & 0.00226004 & 0.03572129 & 0.00031371 & -0.57867667 & 0.14992683 & 28.73966112 & 0.17122002 \\
        StaticVWAP with TKAN & BNB & Volume Curve & 0.10811672 & 0.00083408 & 0.05689332 & 0.00065895 & 0.09560558 & 0.00258979 & 34.58607736 & 3.96786137 \\
        StaticVWAP with LSTM & BNB & Absolute & 0.09585727 & 0.00047720 & 0.03851381 & 0.00097118 & -0.20896325 & 0.04299354 & 10.27560554 & 0.31573055 \\
        StaticVWAP with LSTM & BNB & Quadratic & 0.10212354 & 0.00048888 & 0.03719996 & 0.00024120 & -0.62050756 & 0.01522564 & 9.84294248 & 0.11626344 \\
        StaticVWAP with LSTM & BNB & Volume Curve & 0.10811006 & 0.00061532 & 0.05682895 & 0.00058542 & 0.09668797 & 0.00228171 & 12.07987208 & 1.56730660 \\
        DynamicVWAP with TKAN & BNB & Absolute & 0.08892898 & 0.00011762 & 0.03394928 & 0.00097449 & -0.49685875 & 0.04499367 & 46.82040563 & 4.14006389 \\
        DynamicVWAP with TKAN & BNB & Quadratic & 0.10214637 & 0.00239860 & 0.03297307 & 0.00018456 & -0.71359518 & 0.16002771 & 40.79859676 & 2.44831044 \\
        DynamicVWAP with TKAN & BNB & Volume Curve & 0.10783690 & 0.00150275 & 0.07345307 & 0.00286247 & 0.41924280 & 0.00705824 & 45.43479733 & 3.94142802 \\
        DynamicVWAP with LSTM & BNB & Absolute & \textbf{0.08852550} & 0.00052098 & 0.03474232 & 0.00155376 & -0.51417540 & 0.02936684 & 19.87570839 & 2.12908490 \\
        DynamicVWAP with LSTM & BNB & Quadratic & 0.10123698 & 0.00101242 & \textbf{0.03374521} & 0.00055805 & -0.64225855 & 0.05933125 & 16.25122876 & 0.19874630 \\
        DynamicVWAP with LSTM & BNB & Volume Curve & 0.10719140 & 0.00132662 & 0.07343029 & 0.00208131 & \textbf{0.42767767} & 0.00479149 & 19.53569899 & 0.83384329 \\
        \hline
        Naive & ADA & N/A & 0.15205382 & 0.00000000 & 0.11371138 & 0.00000000 & 0.00000000 & 0.00000000 & 0.00000000 & 0.00000000 \\
        StaticVWAP with TLN & ADA & Absolute & 0.13257912 & 0.00035448 & 0.08265545 & 0.00132954 & -0.19512718 & 0.02112208 & 5.28726745 & 0.20588308 \\
        StaticVWAP with TLN & ADA & Quadratic & 0.13879202 & 0.00074152 & 0.07298782 & 0.00087678 & -0.60492941 & 0.05810657 & 5.19025621 & 0.05141879 \\
        StaticVWAP with TLN & ADA & Volume Curve & 0.14874080 & 0.00053021 & 0.12237117 & 0.00113521 & 0.05929272 & 0.00050809 & 6.14860282 & 0.62042197 \\
        StaticVWAP with TKAN & ADA & Absolute & 0.13082683 & 0.00042942 & 0.08058728 & 0.00155273 & -0.23252373 & 0.04554913 & 30.58913398 & 1.78987412 \\
        StaticVWAP with TKAN & ADA & Quadratic & 0.14050348 & 0.00203601 & 0.07259683 & 0.00053449 & -0.68802873 & 0.06936145 & 28.82713346 & 0.30736096 \\
        StaticVWAP with TKAN & ADA & Volume Curve & 0.14707524 & 0.00143939 & 0.12438853 & 0.00149439 & 0.09960880 & 0.00286178 & 39.45565343 & 4.47713856 \\
        StaticVWAP with LSTM & ADA & Absolute & 0.13072237 & 0.00055947 & 0.08578308 & 0.00203621 & -0.18360337 & 0.01948491 & 11.12554727 & 2.03347702 \\
        StaticVWAP with LSTM & ADA & Quadratic & 0.13841142 & 0.00145406 & 0.07394249 & 0.00105713 & -0.58642588 & 0.05804469 & 9.90328546 & 0.10000470 \\
        StaticVWAP with LSTM & ADA & Volume Curve & 0.14647046 & 0.00028083 & 0.12432448 & 0.00061749 & 0.10201015 & 0.00077251 & 14.52185326 & 2.08953108 \\
        DynamicVWAP with TKAN & ADA & Absolute & 0.12139623 & 0.00099431 & 0.07476030 & 0.00466120 & -0.41400762 & 0.09650446 & 43.64046402 & 3.98394863 \\
        DynamicVWAP with TKAN & ADA & Quadratic & 0.13836165 & 0.00187952 & \textbf{0.06835675} & 0.00090574 & -0.70564603 & 0.07358786 & 39.62808433 & 0.49923827 \\
        DynamicVWAP with TKAN & ADA & Volume Curve & 0.14531860 & 0.00095056 & 0.16246222 & 0.00242535 & 0.43167818 & 0.00338397 & 46.74641275 & 4.34297621 \\
        DynamicVWAP with LSTM & ADA & Absolute & \textbf{0.11911081} & 0.00066513 & 0.07085622 & 0.00212164 & -0.40373142 & 0.05919327 & 18.06567707 & 0.80263690 \\
        DynamicVWAP with LSTM & ADA & Quadratic & 0.13519442 & 0.00153656 & 0.07818373 & 0.00205757 & -0.53143350 & 0.05292493 & 17.78310037 & 2.17485681 \\
        DynamicVWAP with LSTM & ADA & Volume Curve & 0.14607589 & 0.00086770 & 0.16617917 & 0.00205923 & \textbf{0.43936533} & 0.00419243 & 22.92308121 & 2.70775730 \\
        \hline
        Naive & XRP & N/A & 0.14498280 & 0.00000000 & 0.14530325 & 0.00000000 & 0.00000000 & 0.00000000 & 0.00000000 & 0.00000000 \\
        StaticVWAP with TLN & XRP & Absolute & 0.12494882 & 0.00027849 & 0.10233872 & 0.00071432 & -0.25957496 & 0.01187836 & 5.25164285 & 0.15737794 \\
        StaticVWAP with TLN & XRP & Quadratic & 0.13132520 & 0.00161800 & 0.08690822 & 0.00161135 & -0.86853343 & 0.11965111 & 5.21011033 & 0.10212313 \\
        StaticVWAP with TLN & XRP & Volume Curve & 0.14230992 & 0.00059297 & 0.15039715 & 0.00043124 & 0.04383868 & 0.00384918 & 5.66982388 & 0.29291982 \\
        StaticVWAP with TKAN & XRP & Absolute & 0.12393770 & 0.00013537 & 0.10164603 & 0.00263597 & -0.27752671 & 0.03706583 & 32.12862577 & 2.75525176 \\
        StaticVWAP with TKAN & XRP & Quadratic & 0.13296369 & 0.00166776 & 0.08152111 & 0.00118053 & -0.96874272 & 0.09331600 & 30.35712714 & 2.33134681 \\
        StaticVWAP with TKAN & XRP & Volume Curve & 0.14268802 & 0.00149142 & 0.15938525 & 0.00346468 & 0.06559022 & 0.00322346 & 37.26819854 & 4.27918824 \\
        StaticVWAP with LSTM & XRP & Absolute & 0.12417238 & 0.00051768 & 0.10474251 & 0.00264605 & -0.25877017 & 0.05455297 & 10.59912567 & 1.02229319 \\
        StaticVWAP with LSTM & XRP & Quadratic & 0.13249116 & 0.00187386 & 0.08366726 & 0.00150215 & -0.93057966 & 0.09409407 & 9.98833504 & 0.07981146 \\
        StaticVWAP with LSTM & XRP & Volume Curve & 0.14169125 & 0.00098091 & 0.15971367 & 0.00131116 & 0.07089551 & 0.00121417 & 10.80072856 & 0.67601455 \\
        DynamicVWAP with TKAN & XRP & Absolute & 0.11607828 & 0.00058630 & 0.09222955 & 0.00588852 & -0.45512851 & 0.12180167 & 41.96164813 & 2.22023907 \\
        DynamicVWAP with TKAN & XRP & Quadratic & 0.13283433 & 0.00264621 & \textbf{0.07831640} & 0.00132352 & -0.97171995 & 0.10690350 & 41.41955500 & 2.50376753 \\
        DynamicVWAP with TKAN & XRP & Volume Curve & 0.13733773 & 0.00214547 & 0.19822168 & 0.00688708 & 0.41624010 & 0.00585180 & 48.32183676 & 4.44617988 \\
        DynamicVWAP with LSTM & XRP & Absolute & \textbf{0.11426274} & 0.00128449 & 0.08681003 & 0.00333514 & -0.43899483 & 0.11939064 & 17.97093148 & 1.77772746 \\
        DynamicVWAP with LSTM & XRP & Quadratic & 0.13031253 & 0.00143673 & 0.08308718 & 0.00138592 & -0.84227476 & 0.06912297 & 16.86723795 & 0.21748483 \\
        DynamicVWAP with LSTM & XRP & Volume Curve & 0.13772233 & 0.00118413 & 0.19929995 & 0.00265918 & \textbf{0.42127983} & 0.00276343 & 22.56308799 & 3.64669365 \\
        \hline
        \end{tabular}
    }
    \label{table:dynamic_6_step}
\end{table}

\newpage
\newgeometry{left=1cm,right=1cm,top=2cm,bottom=2cm} 

\begin{samepage}
    \subsection{Execution Curves Graphs}
    \label{sec:execution_curves} 
    \vspace{0.5cm}
\end{samepage}

\newcommand{\formatLoss}[1]{%
    \ifnum\pdfstrcmp{#1}{absolute_vwap_loss}=0
        Absolute VWAP Loss%
    \else\ifnum\pdfstrcmp{#1}{quadratic_vwap_loss}=0
        Quadratic VWAP Loss%
    \else\ifnum\pdfstrcmp{#1}{volume_curve_loss}=0
        Volume Curve Loss%
    \fi\fi\fi
}
\captionsetup{font=small}
\newcommand{\assetlist}{BTC,ETH,BNB,ADA,XRP}
\newcommand{\stepslist}{6,12,48}
\newcommand{\losslist}{absolute_vwap_loss, quadratic_vwap_loss, volume_curve_loss}

\foreach \steps in \stepslist {
    \foreach \asset in \assetlist {
        \FloatBarrier
        \ifnum\pdfstrcmp{\steps}{6}=0\relax%
        \ifnum\pdfstrcmp{\asset}{BTC}=0\relax%
        \else
            \clearpage
        \fi
        \else
            \clearpage
        \fi
        
        \begin{samepage}
            \subsubsection{\asset{} Execution Curves with \steps{} Steps Ahead}
            \vspace{0.3cm}
        \end{samepage}
        
        \begin{multicols}{3}
            \foreach \loss in \losslist {
                \begin{figure}[H]
                    \centering
                    \includegraphics[width=0.8\linewidth]
                    {appendix/figures/vwap_execution_allocation_StaticVWAP_with_TLN_using_\loss_\asset_\steps_steps.jpg}
                    \caption{Static w. TLN (\formatLoss{\loss})}
                    \label{fig:dynamic_vwap_static_tln_volume_curve_graph_\asset_\steps_\loss}
                \end{figure}
    
                \begin{figure}[H]
                    \centering
                    \includegraphics[width=0.8\linewidth]
                    {appendix/figures/vwap_execution_allocation_StaticVWAP_with_TKAN_using_\loss_\asset_\steps_steps.jpg}
                    \caption{Static w. TKAN (\formatLoss{\loss})}
                    \label{fig:dynamic_vwap_static_tkan_volume_curve_graph_\asset_\steps_\loss}
                \end{figure}
    
                \begin{figure}[H]
                    \centering
                    \includegraphics[width=0.8\linewidth]
                    {appendix/figures/vwap_execution_allocation_StaticVWAP_with_LSTM_using_\loss_\asset_\steps_steps.jpg}
                    \caption{Static w. LSTM (\formatLoss{\loss})}
                    \label{fig:dynamic_vwap_static_lstm_volume_curve_graph_\asset_\steps_\loss}
                \end{figure}
    
                \begin{figure}[H]
                    \centering
                    \includegraphics[width=0.8\linewidth]
                    {appendix/figures/vwap_execution_allocation_DynamicVWAP_with_TKAN_using_\loss_\asset_\steps_steps.jpg}
                    \caption{Dynamic w. TKAN (\formatLoss{\loss})}
                    \label{fig:dynamic_vwap_dynamic_tkan_volume_curve_graph_\asset_\steps_\loss}
                \end{figure}
    
                \begin{figure}[H]
                    \centering
                    \includegraphics[width=0.8\linewidth]
                    {appendix/figures/vwap_execution_allocation_DynamicVWAP_with_LSTM_using_\loss_\asset_\steps_steps.jpg}
                    \caption{Dynamic w. LSTM (\formatLoss{\loss})}
                    \label{fig:dynamic_vwap_dynamic_lstm_volume_curve_graph_\asset_\steps_\loss}
                \end{figure}
            }

        \end{multicols}
        
        \FloatBarrier
        \clearpage
    }
}
\restoregeometry
\end{appendix}

\end{document}